\documentclass[aps,prc,twocolumn,superscriptaddress,showpacs,floatfix,nofootinbib]{revtex4-1}
\usepackage{amsmath,graphicx,color}

\newcommand{\trento}{T$\mathrel{\protect\raisebox{-2.1pt}{R}}$ENTo}

\begin{document}

\title{Fluid velocity from transverse momentum spectra} 

\author{Anthony Guillen}
\author{Jean-Yves Ollitrault}
\affiliation{Universit\'e Paris Saclay, CNRS, CEA, Institut de physique th\'eorique, F-91191 Gif-sur-Yvette, France} 
%\date{\today}

\begin{abstract}
We parametrize the transverse momentum distribution of outgoing hadrons in ultrarelativistic nucleus-nucleus collisions as a superposition of boosted thermal distributions. 
In this approach, which generalizes the conventional blast wave, the momentum distribution is determined by the distribution of the fluid velocity. 
We analyze the difference between this generalized blast-wave parametrization and a full hydrodynamic calculation.  
We then apply the generalized blast-wave fit to experimental data on Pb+Pb collisions at $\sqrt{s_{\rm NN}}=2.76$~TeV. 
The fit is reasonable up to $p_t\sim 6$~GeV, much beyond the range where hydrodynamics is usually applied, but not perfect. 
Based on the differences between the fit and the data, we argue that an ideal hydrodynamic calculation cannot fit simultaneously all identified particle spectra, irrespective of the specific implementation. 
In particular, data display a significant excess of pions at low $p_t$, whose physical interpretation is discussed.
Data also show that the distribution of the fluid velocity becomes broader as the collision becomes less central. 
This broadening is explained by event-by-event hydrodynamic calculations, where it results from the centrality dependence of initial-state fluctuations.
\end{abstract}
\maketitle

\section{Introduction}

The most fascinating aspect of ultrarelativistic nucleus-nucleus collisions is probably the formation of a tiny fluid droplet~\cite{Busza:2018rrf}, whose temperature is the highest ever achieved in the laboratory~\cite{Gardim:2019xjs}, and which expands collectively before fragmenting into particles. 
It has long been known that this collective dynamics is imprinted into the momentum distributions of outgoing hadrons~\cite{Schnedermann:1993ws}.
This is a less spectacular signature of collectivity than anisotropic flow~\cite{Ollitrault:1992bk,Ackermann:2000tr,Alver:2010gr}, yet a direct and simple one.

The blast-wave parametrization has long been applied~\cite{Siemens:1978pb,Schnedermann:1993ws} to fit experimental data~\cite{Abelev:2008ab,Abelev:2013vea} on transverse momentum distributions. 
It expresses the momentum distribution as a linear superposition of thermal distributions, boosted by the fluid velocity. 
It captures the salient features of the collective dynamics without having to resort to a full hydrodynamic modelization.  
The traditional blast-wave approach uses only a few fit parameters~\cite{Retiere:2003kf}.
We generalize it to an arbitrary linear superposition of boosted thermal distributions. 
This superposition is determined by the distribution of the fluid velocity, which we discretize in practice to limit the dimensionality of parameter space. 

In Sec.~\ref{s:blastwave}, we formulate the generalized blast wave as a well-defined approximation to the freeze-out procedure in an ideal hydrodynamic calculation. 
We dub this approximation ``semi-Cooper--Frye''. 
A byproduct of this formulation is that it provides us with a clear definition of the distribution of the fluid velocity in a hydrodynamic simulation. 
We test the validity of the semi-Cooper--Frye approximation using a full hydrodynamic simulation. 
In Sec.~\ref{s:fittohydro}, we carry out a generalized blast-wave fit to the momentum distributions calculated in ideal hydrodynamics.
We study how the distribution of the fluid velocity extracted from the fit compares with the actual distribution in the hydrodynamic calculation.  
In Sec.~\ref{s:lhcfluid}, we apply the generalized blast-wave fit to experimental data from Pb+Pb collisions at $\sqrt{s_{\rm NN}}=2.76$~TeV. 
We compare results obtained by fitting identified particle spectra, and unidentified, charged particle spectra. 
We study the deviations between the fit and data. 
We argue that these discrepancies are not due to the approximation underlying the blast-wave picture, but correspond to genuine deviations between ideal hydrodynamics and data. 
We discuss their possible interpretation, in particular in terms of dissipative corrections. 
In Sec.~\ref{s:centrality}, we study the centrality dependence of $p_t$ spectra.
We extract the centrality dependence of the fluid velocity distribution from LHC data, and we compare it with the centrality dependence calculated in event-by-event hydrodynamics. 

\section{Ideal hydrodynamics as a sum of boosted thermal distributions}
\label{s:blastwave}

In a hydrodynamic simulation, one assumes that the system formed during the collision quickly equilibrates~\cite{Kurkela:2018wud}. 
The equations of relativistic hydrodynamics are used to model the subsequent expansion into the vacuum until the cohesion of the fluid is lost.
The fluid then fragments into individual hadrons.
We follow the usual simplified procedure where the transition from the fluid phase to free hadrons is an instantaneous freeze out~\cite{Cooper:1974mv,Noronha-Hostler:2013gga,Niemi:2015qia,Kanakubo:2019ogh}, and rescatterings in the hadronic phase~\cite{Teaney:2001av,Petersen:2008dd,Bernhard:2016tnd,Schenke:2019ruo} are neglected. 
Hence, the output of the hydrodynamic calculation is a ``freeze-out hypersurface''~\cite{Kolb:2003dz} from which particles are emitted using the Cooper--Frye procedure~\cite{Cooper:1974mv}. 
In this section, we introduce a simplified treament of freeze-out, dubbed ``semi-Cooper--Frye'', under which the momentum distribution reduces to a superposition of boosted thermal distributions. 
We test its validity using a realistic hydrodynamic calculation. 

\subsection{Why ideal hydrodynamics}
\label{s:whyideal}

Throughout this paper, we use ideal hydrodynamics, not viscous hydrodynamics.
This means in practice that particles are emitted according to a thermal distribution in the rest frame of the fluid~\cite{Ollitrault:2007du}. 
There are two reasons for this choice. 
The first reason is that the validity of hydrodynamics at late times is only guaranteed if the departure from thermal equilibrium is small. 
Hydrodynamics is a gradient expansion~\cite{Baier:2007ix,Romatschke:2017ejr}, where ideal hydrodynamics is the leading term, and Navier--Stokes viscous hydrodynamics represents the first-order 
correction.\footnote{It has been shown that viscous hydrodynamics actually applies beyond the gradient expansion when modeling the boost-invariant  longitudinal expansion at early times~\cite{Heller:2015dha,Romatschke:2017vte}. 
Specifically, it can be applied even when longitudinal momenta are significantly smaller than transverse momenta~\cite{Florkowski:2010cf,Bazow:2013ifa,Blaizot:2017ucy}. 
However, there is no generalization of this result for the generic three-dimensional expansion which applies at late times, and which we consider here. }
While the first-order correction is significant for anisotropic flow~\cite{Romatschke:2007mq}, it is known to be modest for the transverse momentum distributions, averaged over azimuthal angle~\cite{Heinz:2013th}, which we study here.
The second reason is that the deviations from the thermal distribution are not solely determined by the viscosity (shear and bulk).
They depend on the underlying microscopic transport processes~\cite{Dusling:2009df}, such as hadronic interactions at freeze out~\cite{Noronha-Hostler:2013gga,Molnar:2014fva}, which are not constrained.
One of our goals is to obtain direct information on these deviations from experimental data, by studying how data deviate from the ideal-fluid picture.
We come back to this in Sec.~\ref{s:identified}. 

\subsection{Cooper--Frye freeze out}
\label{s:cooperfrye}

The distribution of hadrons at freeze out is 
given by the Cooper--Frye formula~\cite{Cooper:1974mv}:
\begin{equation}
  \label{cooperfrye}
  \frac{dN}{d^3 p}=\frac{2S+1}{(2\pi)^3}\int_\sigma \frac{1}{e^{E^{*}/T_f}\pm 1}\frac{p^\mu}{p^0}     d\sigma_\mu,
\end{equation}
where $p^\mu$ is the four-momentum ($p^0=\sqrt{\vec p^2+m^2}$),
$u^\mu$ is the fluid four-velocity~\cite{Ollitrault:2007du}, $E^{*}\equiv p^\mu u_\mu$ denotes the energy of the hadron in the rest frame of the fluid, $T_f$ is the freeze-out temperature, and the integral runs over the freeze-out hypersurface $\sigma$, whose infinitesimal area vector is $d\sigma_\mu$. 
$2S+1$ denotes the number of independent spin states, and the $+$ and $-$ signs apply to baryons (half-integer $S$) and mesons (integer $S$), respectively.   
We only discuss ultrarelativistic energies, where the baryon chemical potential $\mu_B$ is negligible. 

If freeze out occurs at a constant time, only $d\sigma_0$ is non-vanishing, and the factor $p^\mu/p^0$ is unity. 
Then, the momentum distribution is simply a sum of thermal distributions, boosted by the fluid velocity.
By constrast, for a fluid flowing through a fixed surface, corresponding to the spatial components $d\vec\sigma$,  the thermal distribution is multiplied by the velocity of the particle $\vec p/p^0$. 
Hence, the particle distribution in ideal hydrodynamics is not solely determined by the temperature and the fluid velocity.
It also involves the orientation of the freeze-out hypersurface in space-time. 

\subsection{Semi-Cooper--Frye freeze out} 
\label{s:semicooperfrye}
We introduce a simplified version of the Cooper--Frye formula (\ref{cooperfrye}), where we evaluate the flux by approximating the particle velocity ${\bf p}/p^0$ with the fluid velocity ${\bf u}/u^0$:
\begin{equation}
  \label{cooperfryesimple}
  \frac{dN}{d^3 p}=\frac{2S+1}{(2\pi)^3}\int_\sigma \frac{1}{e^{E^{*}/T_f}\pm 1}\frac{u^\mu}{u^0}     d\sigma_\mu. 
\end{equation}
This expression coincides with (\ref{cooperfrye}) for massive particles in the low-temperature limit, where the thermal velocity is negligible compared to the fluid velocity~\cite{Borghini:2005kd}.
Note that unlike Eq.~(\ref{cooperfrye}), Eq.~(\ref{cooperfryesimple}) is not invariant under Lorentz tranformations.\footnote{Lorentz invariance can be preserved by replacing $p^\mu/p^0$ with $E^* u^\mu/p^0$ in Eq.~(\ref{cooperfrye}). We have checked numerically that this Lorentz-invariant approximation is worse than our semi-Cooper--Frye approximation. The reason is that it is also wrong for the space-like portion of the freeze-out hypersurface, where semi-Cooper--Frye is exact. Note that the Cooper--Frye formula~\cite{Cooper:1974mv} was precisely introduced as a replacement to this Lorentz-invariant approximation~\cite{Hagedorn:1967tlw}.}
Lorentz invariance is however irrelevant here because Eq.~(\ref{cooperfryesimple}) is merely an approximation which we use only in a specific reference frame, which will be defined below. 
We dub the approximation (\ref{cooperfryesimple}) ``semi-Cooper--Frye'' freeze out because it coincides with the Cooper--Frye formula only for the space-like part of the hypersurface.\footnote{Note that the semi-Cooper--Frye ansatz automatically solves the problem of negative contributions in Eq.~(\ref{cooperfrye})~\cite{Bugaev:2002ch,Oliinychenko:2014tqa}, because the fluid typically flows outwards, at least for smooth initial conditions. The price to pay is a slight violation of energy and momentum conservation.}
Its validity will be tested in Sec.~\ref{s:semitest}. 

The distribution (\ref{cooperfryesimple})  can be rewritten as an integral over the fluid velocity ${\bf u}=(u_x,u_y,u_z)$:
\begin{equation}
 \label{cooperfryeuxyz}
  \frac{dN}{d^3 p}=\frac{2S+1}{(2\pi)^3}\int \frac{1}{e^{E^{*}/T_f}\pm 1}
\Omega({\bf u}) d{\bf u},
\end{equation}
where $\Omega({\bf u})d{\bf u}$ is defined as: 
\begin{equation}
  \label{probauxyz}
 \Omega({\bf u}) d{\bf u}= \int_{\sigma,{\bf u}\ {\rm in}\ d{\bf u}} \frac{u^\mu}{u^0}      d\sigma_\mu. 
\end{equation}
In Eq.~(\ref{probauxyz}), the integral runs only on the part of the freeze-out hypersurface $\sigma$ where the fluid velocity is ${\bf u}$ up to $d{\bf u}$. 
If freeze-out occurs at a constant time,  only the $\mu=0$ component contributes, and $\Omega({\bf u})d{\bf u}$ is simply the volume of the fluid with velocity ${\bf u}$ up to $d{\bf u}$. 
The more general expression (\ref{probauxyz}) can be thought of as an effective volume. 

Equation~(\ref{cooperfryeuxyz}) expresses the momentum distribution as a weighted sum of boosted thermal distributions,  and the information about the freeze-out hypersurface is encoded in the weight $\Omega({\bf u})$.  
Hence, the semi-Cooper--Frye approximation can be seen as a generalization of the blast-wave parametrization~\cite{Siemens:1978pb,Mazeliauskas:2019ifr}. 
In the blast-wave parametrization, one assumes that the fluid velocity is some specific function (typically a power law) of the spatial coordinate at freeze-out~\cite{Retiere:2003kf}.\footnote{Note that blast-wave calculations can be improved to take into account the shape of the freeze-out hypersurface~\cite{Yang:2020oig}.}
This amounts to specifying the functional form of $\Omega({\bf u})$. 

A byproduct of our formulation is that Eq.~(\ref{probauxyz}) defines the distribution of the fluid velocity, $\Omega({\bf u})$, as an integral over the freeze-out surface.  

\subsection{Adding decays}
\label{s:fastreso}

Unstable hadrons decay before reaching the detectors, and the measured momentum distributions are those of the decay products at the end of the decay chain. 
We evaluate the momentum distribution of these decay products using the method recently introduced by Mazeliauskas {\it et al.\/}~\cite{Mazeliauskas:2018irt}. 
It amounts to carrying out the following replacement in the Cooper--Frye formula (\ref{cooperfrye}): 
\begin{equation}
\label{aleksastrick}
\frac{1}{e^{E^{*}/T_f}\pm 1} \frac{p^\mu}{p^0}
 \rightarrow f_1(E^{*})\frac{p^\mu}{p^0}+\left(f_2(E^{*})-f_1(E^{*})\right)\frac{E^{*} u^\mu}{p^0},
\end{equation}
where $f_1(E^{*})$ and $f_2(E^{*})$ are two Lorentz scalar functions which are computed for each stable hadron, and depend on the freeze-out temperature $T_f$. They take into account the whole decay chain. 
If there are no decays, then $f_2(E^{*})=f_1(E^{*})=(e^{E^{*}/T_f}\pm 1)^{-1}$, and the left-hand side and right-hand side of (\ref{aleksastrick}) coincide. 

Semi-Cooper--Frye freeze out can be readily generalized to include these decays. 
We again approximate the particle velocity $p^\mu/p^0$ with the fluid velocity $u^\mu/u^0$ in the first term of the right-hand side of (\ref{aleksastrick}). 
This amounts to replacing 
 \begin{equation}
\label{aleksastrick0}
\frac{1}{e^{E^{*}/T_f}\pm 1}
 \rightarrow f_1(E^{*})+\left(f_2(E^{*})-f_1(E^{*})\right)
\frac{E^{*} u^0}{p^0}
\end{equation}
 in Eqs.~(\ref{cooperfryesimple}) and  (\ref{cooperfryeuxyz}).
The expression of the effective volume (\ref{probauxyz}) is unchanged. 

\begin{figure*}[ht]
\begin{center}
\includegraphics[scale=1.]{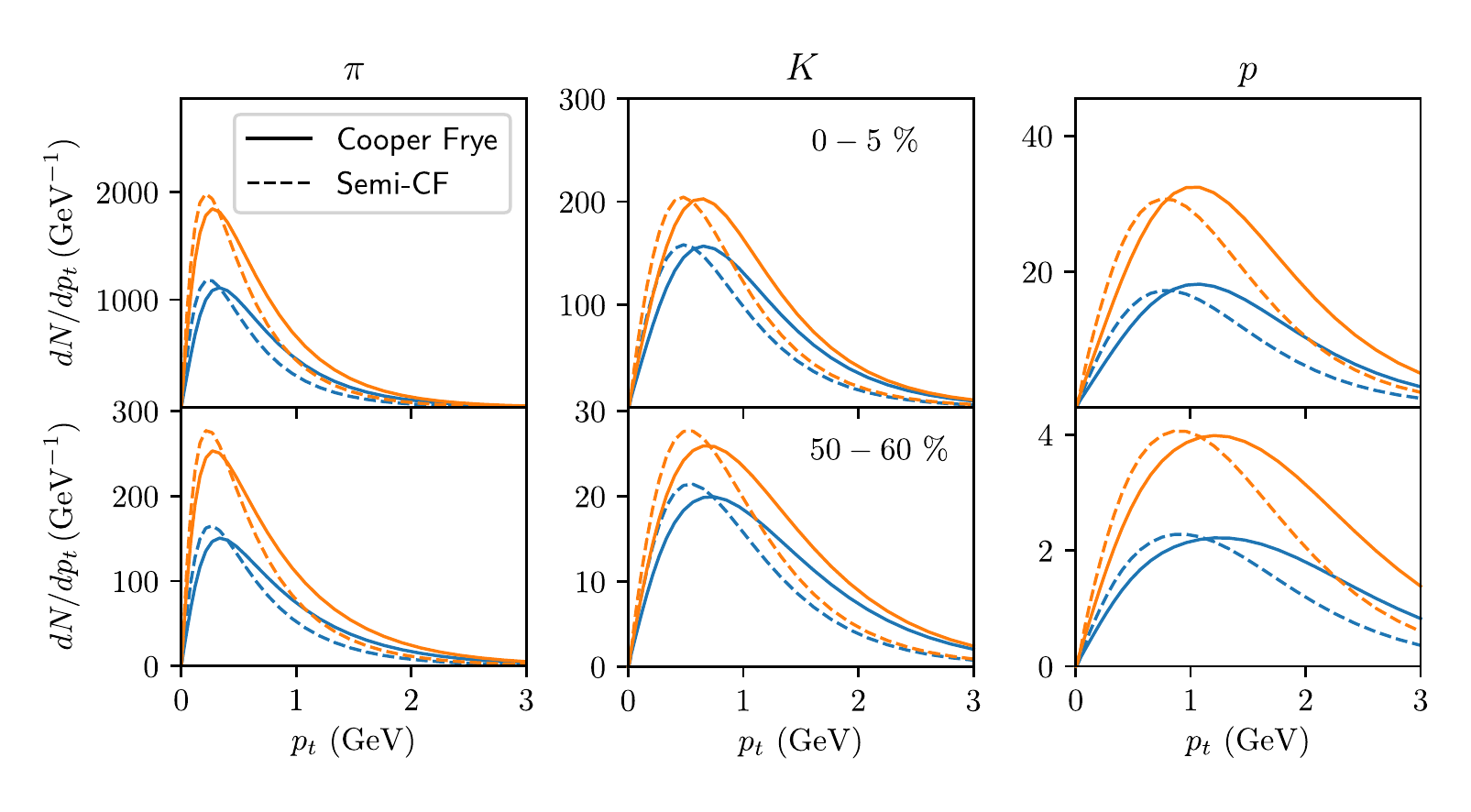} 
\end{center}
\caption{(Color online)
Comparison between $p_t$ spectra calculated using the usual Cooper--Frye formula (\ref{cooperfrye}) (full lines), and the semi-Cooper--Frye approximation (\ref{cooperfryesimple}) (dashed lines), before (blue lines) and after decays (orange lines).
 From left to right: 
Charged pions, 
charged kaons, 
protons plus antiprotons.  
Two random Pb+Pb collisions at $\sqrt{s_{\rm NN}}=2.76$~TeV are shown: a central collision ($b=2.88$~fm, falling in the 0-5\% centrality window) in the top panel, and a more peripheral collision ($b=10.89$~fm, corresponding to a centrality  close to 50\%) in the bottom panel. 
These collisions are modeled using an ideal hydrodynamic simulation which includes initial-state fluctuations, and whose details are given in Sec.~\ref{s:omegau}. 
The freeze-out temperature is $T_f=135$~MeV (see Sec.~\ref{s:freezeout}).
}
\label{fig:cooperfrye}
\end{figure*}   

\subsection{Transverse momentum distribution}
\label{s:ptdis}

From now on, our study is restricted to the transverse momentum distribution, $dN/dp_t$, which is obtained by integrating over the longitudinal momentum and the azimuthal angle:
\begin{equation}
\label{intpzphi}
\frac{dN}{dp_t}=p_t\int_{-\infty}^{+\infty} dp_z \int_{-\pi}^{\pi}d\phi\, \frac{dN}{d^3p}.
\end{equation} 
We now evaluate $dN/dp_t$ using the semi-Cooper--Frye approximation.
$p_t$ is invariant under longitudinal Lorentz boosts, but the semi-Cooper--Frye approximation is not.
Therefore, we need to specify the frame where it is applied.
We choose the reference frame where $u_z=0$, i.e., the fluid is at midrapidity.\footnote{Since the fluid typically extends over a range of rapidities, this implies in practice that we slice the fluid according to rapidity, and evaluate $dN/dp_t$ separately in each slice.}
Then, the momentum distribution is given by Eq.~(\ref{cooperfryeuxyz}), where 
\begin{equation}
  \label{defbare}
  E^{*}\equiv \sqrt{(1+u^2)(m^2+p_z^2+p_t^2)}-up_t\cos(\phi-\phi_u).
\end{equation}
In this equation, $\phi_u$ denotes the azimuthal angle of the fluid velocity, and $u$ the transverse fluid velocity: 
\begin{equation}
  \label{defu}
  u\equiv \sqrt{u_x^2+u_y^2}. 
\end{equation}
Since the $p_t$ distribution (\ref{intpzphi}) is integrated over $\phi$, one can set $\phi_u=0$ in Eq.~(\ref{defbare}) without any loss of generality.
Therefore, the $p_t$ distribution only involves the distribution of the transverse velocity $u$. 

Inserting Eqs.~(\ref{cooperfryeuxyz}) and (\ref{defbare}) into Eq.(\ref{intpzphi}), we thus rewrite the $p_t$ distribution as integral over $u$:
\begin{equation}
  \label{boostedthermal}
\frac{dN}{dp_t}=\int_0^{\infty} f(p_t,u) \Omega(u) du, 
\end{equation}
where $\Omega(u)du$ represents the effective freeze-out volume whose transverse velocity lies between $u$ and $u+du$, obtained by integrating Eq.~(\ref{probauxyz}) on $u_z$ and $\phi_u$: 
\begin{equation}
  \label{probau}
 \Omega(u) du= \int_{\sigma, u\ {\rm in}\ du} \frac{u^\mu}{\sqrt{1+u^2}}      d\sigma_\mu
\end{equation}
(note that the $\mu=z$ component is zero by choice of the reference frame), 
and $f(p_t,u)$ is a boosted thermal distribution: 
\begin{equation}
  \label{deffptu}
f(p_t,u)\equiv \frac{2S+1}{(2\pi)^3}
p_t\int_{-\infty}^{+\infty}dp_z\int_{-\pi}^{\pi}d\phi \frac{1}{e^{E^{*}/T_f}\pm 1}. 
\end{equation}
Resonance decays are taken into account through the substitution (\ref{aleksastrick0}).

Equations~(\ref{boostedthermal}) and (\ref{probau}) are simplifications of the more general Eqs.~(\ref{cooperfryeuxyz}) and (\ref{probauxyz}), in the sense that they are restricted to the transverse components of the momentum, and of the fluid velocity.   

\subsection{Testing the semi-Cooper--Frye approximation}
\label{s:semitest}

We test the validity of the semi-Cooper--Frye approximation by comparing with the results of the standard Cooper--Frye procedure for an ideal hydrodynamic simulation of a central Pb+Pb collision at $\sqrt{s_{\rm NN}}=2.76$~TeV. 
Our simulation, whose details will be specified below in Sec.~\ref{s:omegau}, assumes longitudinal boost invariance~\cite{Bjorken:1982qr}. 
Hence, the hypersurface element $d\sigma_\mu$ must be understood as ``per unit rapidity'', and so is the resulting momentum distribution.
Therefore, $dN/dp_t$ in (\ref{boostedthermal}) actually stands for $dN/dydp_t$, where $y$ is the rapidity.
Resonance decays are implemented using the FastReso code~\cite{Mazeliauskas:2018irt} with the list of resonances from Ref.~\cite{Alba:2020jir}.
The semi-Cooper--Frye approximation is implemented by first evaluating the distribution of the transverse fluid velocity at freeze-out using Eq.~(\ref{probau}), and then computing the spectra using Eq.~(\ref{boostedthermal}). 

The comparison with the usual Cooper--Frye result is displayed in Fig.~\ref{fig:cooperfrye}, for a central event (top panel) and for a more peripheral event (bottom panel). 
The semi-Cooper--Frye approximation captures the main features of the $p_t$ spectra, namely, the overall shape and scale, and the mass ordering. 
However, it overestimates the particle yield at low $p_t$ and underestimates it at high $p_t$, compared to the full Cooper--Frye treatment. 
This can be readily understood by comparing the corresponding equations: Eq.~(\ref{cooperfryesimple})  overestimates or underestimates the particle yield, relative to Eq.~(\ref{cooperfrye}), depending on whether the fluid velocity $\vec u/u^0$ is larger or smaller than the particle velocity $\vec p/p^0$. 
The semi-Cooper--Frye approximation underestimates both the particle yields, obtained by integrating the spectra (by 14\%, 16\%, 16\% for pions, kaons, protons for the top panel of Fig.~\ref{fig:cooperfrye}), and the mean transverse momentum per particle (by 17\%, 17\%, 14\% for pions, kaons and protons). 
These numbers provide us with a quantitative estimate of the violation of energy-momentum conservation induced by the semi-Cooper--Frye approximation. 
The dominant contribution to the energy comes from the pions, and the energy of a pion is close to its transverse momentum. 
The total energy, obtained by multipying the multiplicity with the mean transverse momentum, is underestimated by almost 30\%. 
Note that the semi-Cooper--Frye approximation is not worse for peripheral collisions than for central collisions, as shown by the bottom panel of Fig.~\ref{fig:cooperfrye}. 

Despite these differences, one should keep in mind that spectra are usually shown on a logarithmic scale. 
As we shall see in Sec.~\ref{s:lhcfluid}, the discrepancies between hydrodynamics and experimental data are typically larger than the error introduced by the semi-Cooper--Frye approximation, so that a generalized blast-wave fit is a decent approximation to a full hydrodynamic calculation. 

\begin{figure}[ht]
\begin{center}
\includegraphics[width=\linewidth]{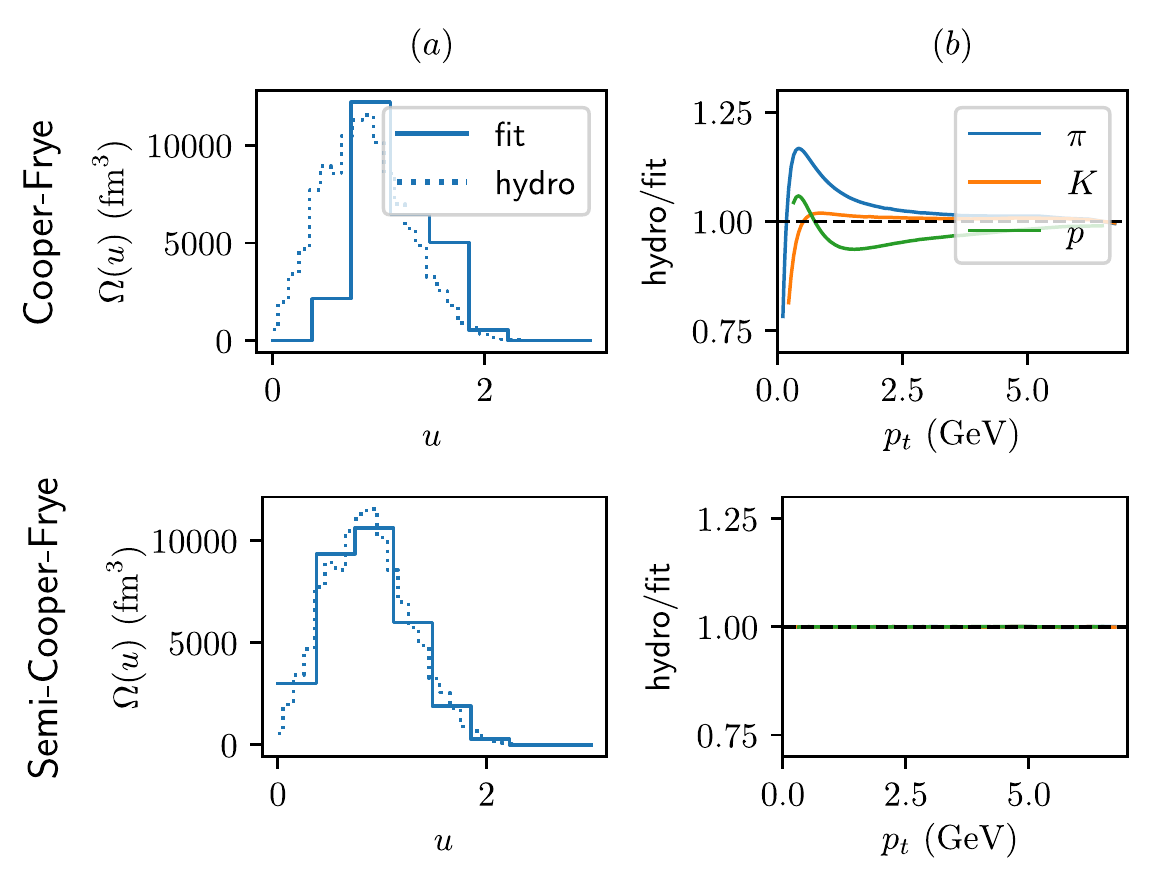} 
\end{center}
\caption{(Color online) 
Left: distribution of the fluid velocity $\Omega(u)$ corresponding to the hydrodynamic calculation of Fig.~\ref{fig:cooperfrye}. 
The dotted line is the value of $\Omega(u)$ calculated directly from the freeze-out surface using Eq.~(\ref{probau}). 
The step functions are the values obtained from a combined fit of the pion, kaon and proton $dN/dp_t$ using Eq.~(\ref{boostedthermal}), and the right panels display the ratio of the ``data'' ($dN/dp_t$ from the hydrodynamic calculation)  to the generalized blast-wave fit.  
Top: results obtained by fitting the Cooper--Frye spectra (full lines in Fig.~\ref{fig:cooperfrye}).
Bottom: results obtained by fitting the semi-Cooper--Frye spectra (dashed lines in Fig.~\ref{fig:cooperfrye}).
}
\label{fig:fittohydro}
\end{figure}

\section{Generalized blast-wave fit to hydrodynamics} 
\label{s:fittohydro}

A generalized blast-wave fit to the momentum distribution, $dN/dp_t$, using Eq.~(\ref{boostedthermal}) returns the preferred distribution of the fluid velocity, $\Omega(u)$. 
This procedure can be applied not only to the experimental $dN/dp_t$, but also to the result of a hydrodynamic simulation. 
In this section, we study how $\Omega(u)$ from a generalized blast-wave fit compares with that obtained by a direct integration over the freeze-out surface using Eq.~(\ref{probau}). 
We use the same hydrodynamic calculation as in Sec.~\ref{s:semitest}, corresponding to a central Pb+Pb collision at $\sqrt{s_{\rm NN}}=2.76$~TeV. 
The distribution of the fluid velocity defined by Eq.~(\ref{probau}) is displayed as a dotted line in the two left panels of Fig.~\ref{fig:fittohydro}. 

We first carry out a consistency check in the following way. 
We calculate $dN/p_t$ using Eq.~(\ref{boostedthermal}), with $\Omega(u)$ from Eq.~(\ref{probau}) as an input.
These distributions correspond to the dashed lines in Fig.~\ref{fig:cooperfrye}.
We then reconstruct $\Omega(u)$ by fitting $dN/p_t$ using the same equation (\ref{boostedthermal}), and the same temperature. 
The value of $\Omega(u)$ returned by the fit should be identical to the input value. 
We carry out a combined fit of pion, kaon and proton $dN/dp_t$ up to $p_t=7$~GeV/c. 
In order to limit the number of fit parameters, we have assumed that $u$ takes discrete values spaced with a step $\Delta u=0.4$. 
The corresponding $\Omega(u)$ is a sum of Dirac peaks centered at the corresponding values of $u$, which we represent in the bottom left panel of Fig.~\ref{fig:fittohydro} as a step function with the same area. 
The reconstructed $\Omega(u)$ matches with the input value, up to the discretization. 
Despite the discretization, the fit is essentially perfect, as shown by the ratio displayed in the bottom right panel. 
This consistency check validates our fitting algorithm. 
More importantly, it illustrates the level of detail to which one can hope to reconstruct the fluid velocity distribution $\Omega(u)$. 

We now apply the same fitting procedure to the spectra from the full hydrodynamic calculation, corresponding to the full lines in Fig.~\ref{fig:cooperfrye}. 
The resulting $\Omega(u)$ is displayed in the top left panel of Fig.~\ref{fig:fittohydro}. 
One notices several differences with the input value. 
First, the distribution is shifted to the right. 
This shows that the generalized blast-wave fit somewhat overestimates the fluid velocity. 
This compensates the fact that the semi-Cooper--Frye approximation (which underlies the blast-wave picture) underestimates the mean transverse momentum (dashed lines in Fig.~\ref{fig:cooperfrye}). 
Second, the reconstructed $\Omega(u)$ decreases much faster for small $u$ than the input distribution. 
The reason is that the semi-Cooper--Frye approximation largely overestimates the particle yield at low $p_t$ (compare the dashed lines and the full lines in Fig.~\ref{fig:cooperfrye}). 
The fit partially compensates for this effect by suppressing the low values of the fluid velocity $u$, which are the dominant sources of low-$p_t$ particles. 
The compensation is only partial, as the blast-wave fit still overestimates the particle yield at extremely low $p_t$ (right panel in Fig.~\ref{fig:fittohydro}). 
Because of this strong suppression at small $u$, the reconstructed $\Omega(u)$ is significantly narrower than the input. 
We come back to this when we discuss experimental data in Sec.~\ref{s:lhcfluid}.

\section{Generalized blast-wave fits to LHC data} 
\label{s:lhcfluid}

We now apply the generalized blast-wave fit to LHC data and extract the distribution of the fluid velocity $\Omega(u)$ from the measured spectra. 
The choice of the freeze-out temperature $T_f$ is discussed in Sec.~\ref{s:freezeout}.  
We carry out combined fits to pion, kaon, proton spectra in Sec.~\ref{s:identified}. 
In Sec.~\ref{s:omegau}, the values of $\Omega(u)$ from the fits are compared with those from hydrodynamic calculations. 
Fits to unidentified hadron spectra are discussed in Sec.~\ref{s:unidentified}. 

\subsection{Freeze-out temperature}
\label{s:freezeout}

\begin{figure}[h]
\begin{center}
\includegraphics[scale=1.]{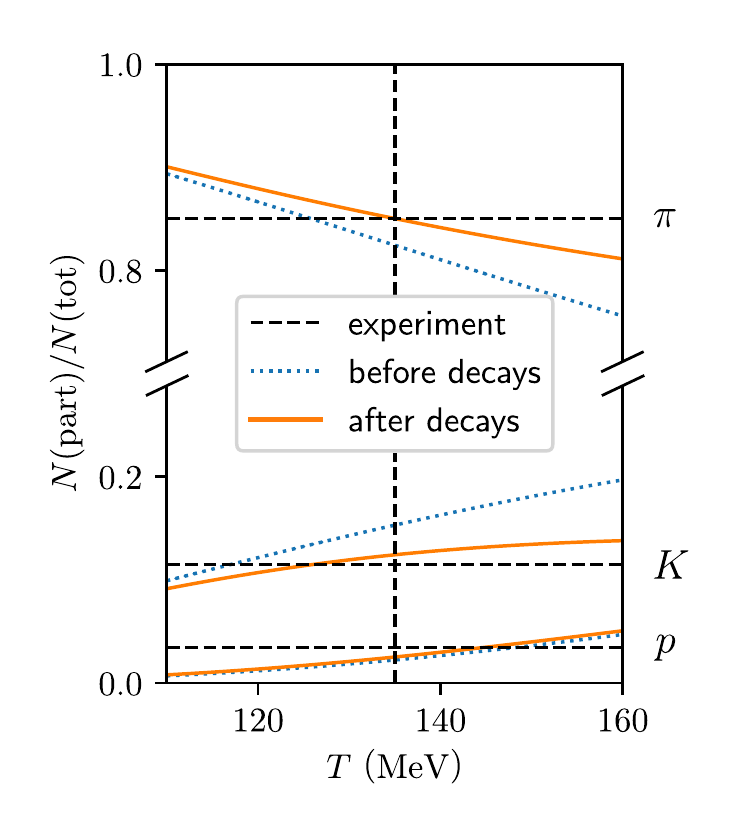} 
\end{center}
\caption{(Color online)
  Relative abundances of charged pions, kaons, and protons, before (dotted lines) and after (full line) resonance decays~\cite{Mazeliauskas:2018irt}  for a fluid at rest, as a function of its temperature $T_f$.
  These relative abundances add up to unity by construction.
  The horizontal dashed lines correspond to the experimental values, obtained by integrating the spectra in 0-5\% central Pb+Pb collisions at $\sqrt{s_{\rm NN}}=2.76$~TeV~\cite{Adam:2015kca}. 
  The range of the integration over $p_t$ in our calculation is the same as in experiment. 
  The vertical line corresponds to the value $T_f=135$~MeV which we choose throughout this article. 
}
\label{fig:tf}
\end{figure}    

We consider for simplicity a single freeze-out model~\cite{Broniowski:2001uk,Noronha-Hostler:2013gga} where chemical and kinetic freeze-out occur simultaneously at temperature $T_f$.
In other terms, we do not implement partial chemical equilibration~\cite{Bebie:1991ij,Huovinen:2007xh,Niemi:2015qia}.
Equation~(\ref{boostedthermal}) then defines not only the probability distribution of $p_t$ for a given hadron, but also its normalization. 
Since we are going to fit the spectra of pions, kaons and protons~\cite{Abelev:2013vea,Adam:2015kca}, we must choose a value of $T_f$ which fits their relative abundances, obtained as ratios of integrated spectra.
Figure~\ref{fig:tf} displays the relative abundances of these particles as a function of $T_f$ for a fluid at rest. 
The choice $T_f=135$~MeV gives reasonable agreement with experiment, once the feed-down from resonance decays is taken into account. 
We choose this value throughout this article.\footnote{Note that the relative abundances are strictly independent of the fluid velocity if integrated over all $p_t$. In Fig.~\ref{fig:tf}, however, we implement the same $p_t$ cuts as in experiment. Therefore, the relative abundances are no longer strictly independent of the fluid velocity, but we neglect this dependence.}
Devetak {\it et al.\/}~\cite{Devetak:2019lsk} obtain a similar value (137~MeV) through a global fit to pion, kaon, proton spectra.\footnote{Note that relative yields are somewhat modified if one takes into account finite resonance widths~\cite{Vovchenko:2018fmh} and pion-nucleon interactions~\cite{Andronic:2018qqt}, which are both neglected here.}

Note that our value of $T_f$ is significantly smaller than the usual value $T_c=156$~MeV of the chemical freeze-out temperature~\cite{Andronic:2017pug}, obtained by fitting the relative abundances of {\it all\/} hadrons.
This higher temperature is mostly dictated by relative abundances of strange baryons~\cite{Alba:2020jir}.
But we do not study strange baryons here, and they represent a small fraction of the particle yield anyway (see Sec.~\ref{s:unidentified} for the case of $\Sigma$ baryons). 

Note also that our approach differs from usual blast-wave fits, where the freeze-out temperature is fitted independently for each hadron species~\cite{Abelev:2008ab,Abelev:2013vea}.
Our goal is to mimic a hydrodynamic calculation, where the freeze-out temperature is common to all hadrons, and determines both the spectra and the relative yields. 

\subsection{Fits to identified particle spectra} 
\label{s:identified}

\begin{figure*}[ht]
\begin{center}
\includegraphics[scale=1.]{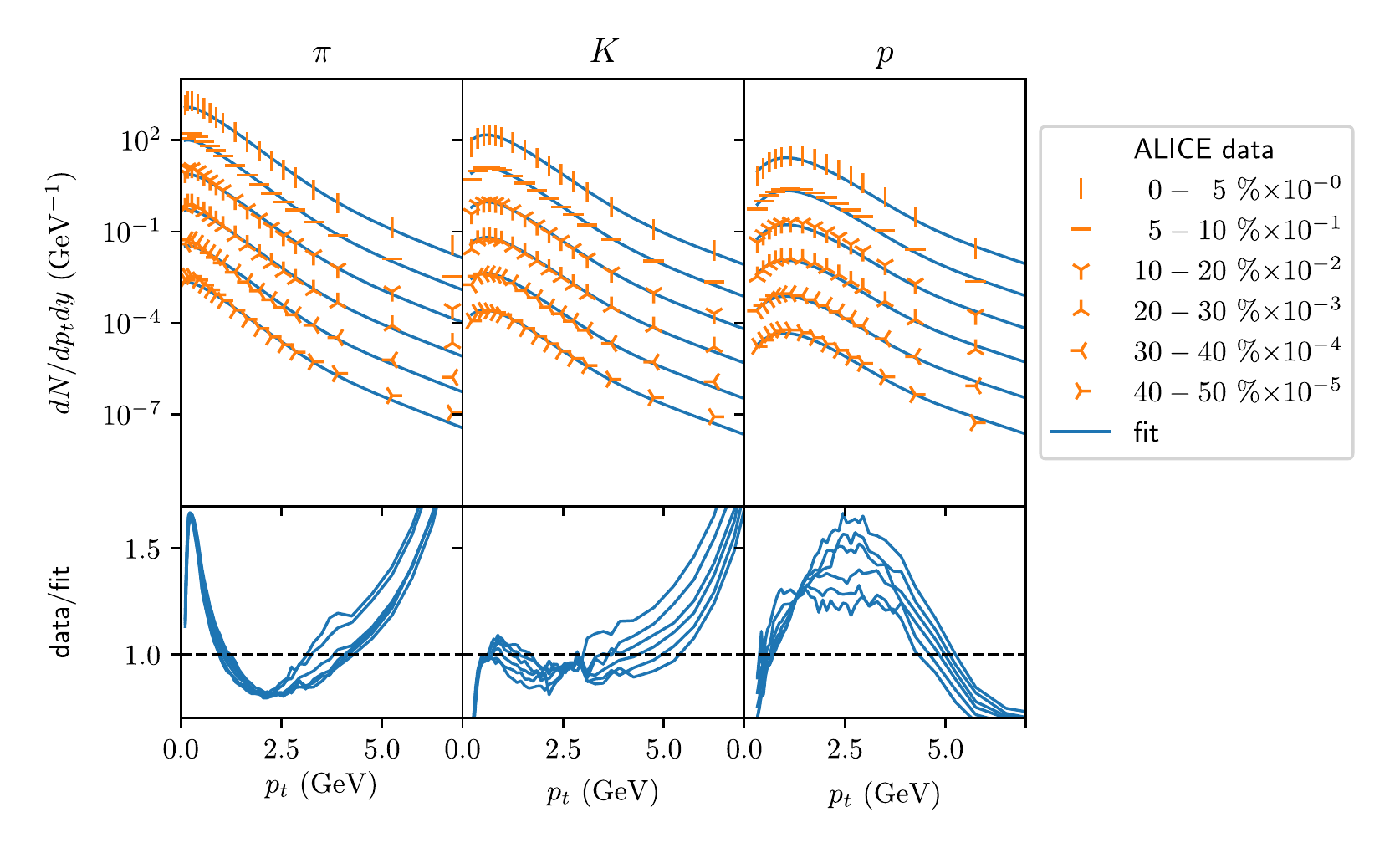} 
\end{center}
\caption{(Color online)
  Symbols: $p_t$ distributions of charged pions, charged kaons and  protons measured by ALICE~\cite{Adam:2015kca} in Pb+Pb collisions at $\sqrt{s_{\rm NN}}=2.76$~TeV, in several centrality windows. 
  Lines: Fits using Eq.~(\ref{boostedthermal}). 
  For each centrality, the function $\Omega(u)$ is fitted so as to achieve the best simultaneous fit of pion, kaon, proton spectra in the range $0<p_t<7$~GeV/c. 
  The bottom panels display the ratio data/fit. 
}
\label{fig:identifiedspectra}
\end{figure*} 

We carry out a combined fit of pion, kaon and proton spectra measured in Pb+Pb collisions at $\sqrt{s_{\rm NN}}=2.76$~TeV~\cite{Adam:2015kca} up to $p_t=7$~GeV/c using Eq.~(\ref{boostedthermal}), following the exact same procedure as in Sec.~\ref{s:fittohydro}. 
The procedure is repeated in every centrality window. 
Fig.~\ref{fig:identifiedspectra} displays the measured spectra together with the fits.
Data are above the fits  at high $p_t$, as expected for thermal models.
Note, however, that the deviations become large only for $p_t>6$~GeV. 
This is much higher than the typical $p_t$ range used in hydrodynamic calculations~\cite{Ryu:2017qzn,Dubla:2018czx,Devetak:2019lsk,Everett:2020xug}  or blast-wave fits~\cite{Mazeliauskas:2019ifr,Melo:2019mpn}, which typically do not extend beyond $p_t\sim 2-3$~GeV.\footnote{The $p_t$ range can be extended by replacing the thermal distribution with a distribution that decreases more slowly at large momentum. This is the so-called ``Tsallis-blast-wave'' fit approach~\cite{Tang:2008ud}.}
It is interesting to note that our simple generalization of the blast-wave approach allows us to greatly improve agreement with data all the way to $6$~GeV. 
The reason will be discussed in Sec.~\ref{s:omegau}. 

Deviations between the fit and data also appear at lower transverse momentum. 
Data show an excess of pions for $p_t<1$~GeV/c, followed by a depletion up to $p_t\sim 3$~GeV, and an excess of protons for $p_t>1$~GeV/c. 
Similar deviations have been reported by other authors~\cite{Melo:2019mpn,Mazeliauskas:2019ifr}. 
The first question is whether data would be in better agreement with a full ideal hydrodynamic calculation, than with the generalized blast-wave fit.  
As shown in the top right panel of  Fig.~\ref{fig:fittohydro}, the differences betwen the full hydrodynamic calculation and a blast-wave fit are only significant at low $p_t$. 
In particular, the hydrodynamic calculation shows a pion excess relative to the fit for $p_t<0.5$~GeV/c, but this excess is much smaller than that of data relative to the fit. 
We therefore conclude that there would still be a pion excess at low $p_t$~\cite{Alqahtani:2017tnq,Dubla:2018czx} if one compared experimental data with a full ideal hydrodynamic calculation, irrespective of the details of this calculation. 
The excess is present even though we took into account the feed-down from resonance decays, whose contribution is essential at low $p_t$. 

There are two possible explanations for this excess. 
The simplest explanation is to attribute it to the leading correction to ideal hydrodynamics, namely, viscous hydrodynamics. 
As recalled in Sec.~\ref{s:whyideal}, the viscous correction to the thermal distribution is not universal. 
It depends on momentum through details of hadron cross sections. 
Essentially all viscous hydrodynamic calculations~\cite{Ryu:2017qzn,Dubla:2018czx,Devetak:2019lsk,Everett:2020xug}
 assume for simplicity that the momentum dependence is quadratic~\cite{Teaney:2003kp}, which in turn implies that the departure from thermal equilibrium is larger at high $p_t$. 
This quadratic ansatz makes hydrodynamic calculations of anisotropic flow look better at high $p_t$~\cite{Heinz:2013th}, but lacks a microscopic justification~\cite{Dusling:2009df,Molnar:2014fva}. 
Our comparison to data suggests instead that the departure from thermal equilibrium is larger for low momentum pions. 
This seems natural from a theoretical point of view, since low-momentum pions are Goldstone bosons~\cite{Gasser:1983yg,Colangelo:2001df} which interact little.

A more radical scenario is that low-momentum pions interact so little that should be treated as a separate, superfluid-like, component~\cite{Grossi:2020ezz}. 
It has been recently argued that this scenario may explain the observed pion excess at low $p_t$~\cite{Grossi:2021gqi}. 

\subsection{Distribution of fluid velocity}
\label{s:omegau}
\begin{figure}[ht]
\begin{center}
\includegraphics[width=\linewidth]{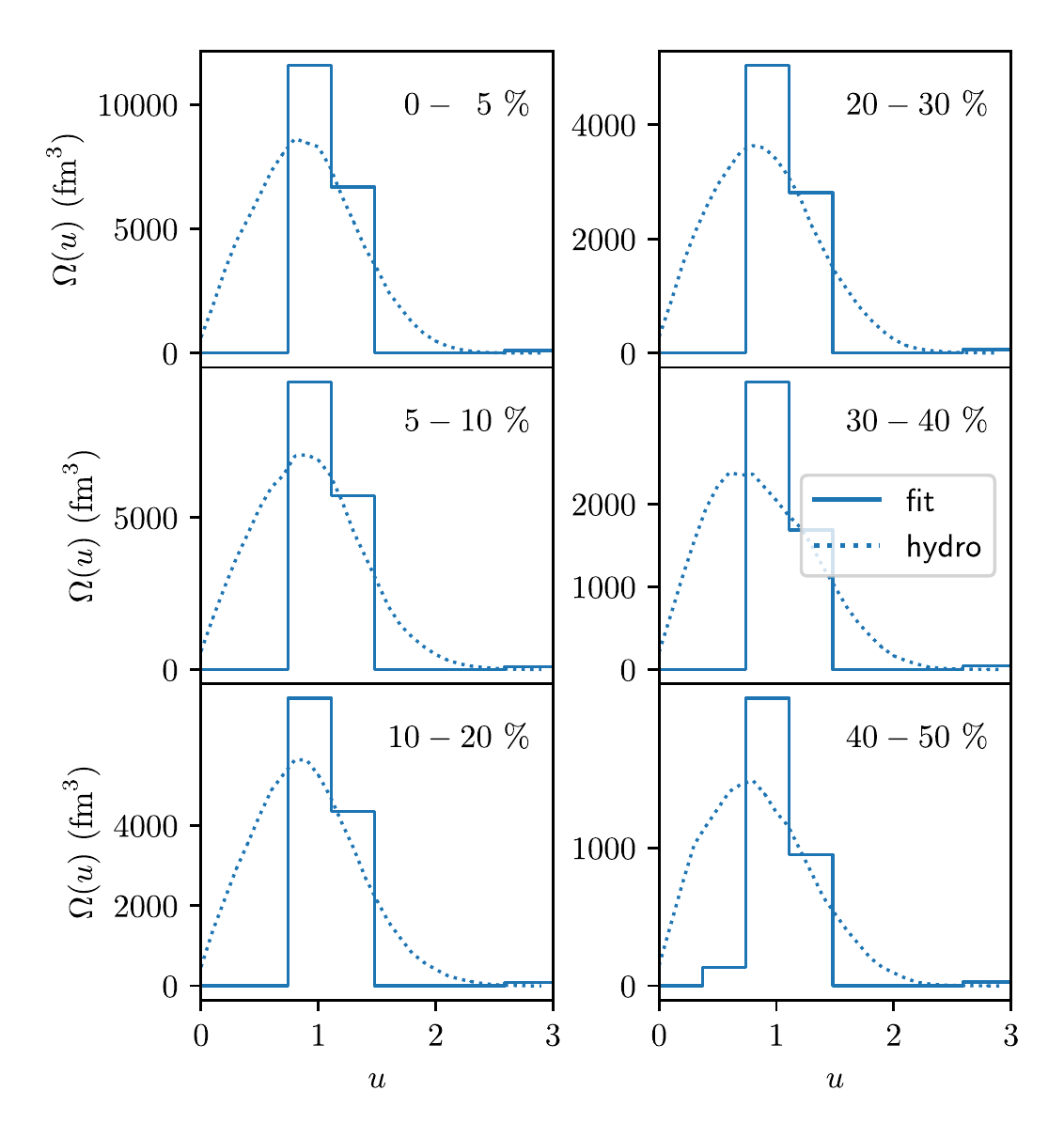} 
\end{center}
\caption{(Color online)
  Full lines:
Values of $\Omega(u)$ given by the fits of Fig.~\ref{fig:identifiedspectra}.  
As in Fig.~\ref{fig:fittohydro}, we have assumed that $u$ takes discrete values with a step $\Delta u=0.4$. 
  The integral $V_{\rm eff}\equiv\int_{u=0}^{\infty} \Omega(u)du$ is an effective volume per unit rapidity at freeze-out, hence the unit fm$^3$ on the vertical axis. 
  Dotted curves: $\Omega(u)$ in an event-by-event ideal hydrodynamic calculation with \trento{} initial conditions (see text). 
}
\label{fig:Omegaidentified}
\end{figure}

We now discuss the results for the fitting function $\Omega(u)$, whose value is returned by the fits to identified particle spectra.
As in Sec.~\ref{s:fittohydro}, we have assumed that $u$ takes discrete values spaced with a step $\Delta u=0.4$. 
The values of $\Omega(u)$ given by the fits of Fig.~\ref{fig:identifiedspectra} are represented in Fig.~\ref{fig:Omegaidentified}.
The most probable values of $u$ are around unity, corresponding to a fluid moving at 70\% of the velocity of light. 
The integral $V_{\rm eff}\equiv\int\Omega(u)du$ is the effective volume at freeze out. 
It represents the volume of a hadron gas at $T_f=135$~MeV such that the multiplicity is the same as in data. 
Therefore, its centrality dependence follows that of the charged multiplicity~\cite{Aamodt:2010cz}. 

Note also that $\Omega(u)$ has a non-zero value in the highest velocity bin $2.6<u<3$, corresponding to collective velocities in the range $93-95$\% of the velocity of light. 
$\Omega(u)$ in this bin is very small, yet the corresponding contribution becomes dominant at high $p_t$.  
This is the ``trick'' that enables the fitting algorithm to fit data all the way up to $p_t\sim 6$~GeV.
Usual blast-wave fits use a smooth $\Omega(u)$, whose support (the range in $u$ where $\Omega(u)>0$) is much smaller, and fail typically beyond 2~GeV. 
A lump of fluid moving at a velocity close to the velocity of light is the equivalent, within a hydrodynamic description, of a jet. 
Our result suggests that the inclusion of minijets~\cite{Paatelainen:2013eea} in full hydrodynamic simulations may help extend their validity to higher $p_t$. 

For the sake of comparison with the result of our fit, we now evaluate $\Omega(u)$ in a state-of-the-art ideal hydrodynamic simulation of Pb+Pb collisions at $\sqrt{s_{\rm NN}}=2.76$~TeV, which we now describe. 
We use boost-invariant~\cite{Bjorken:1982qr} initial conditions, with a starting time $\tau_0=0.4$~fm/c. 
The transverse velocity at $\tau_0$ is set to zero, that is, initial flow~\cite{Vredevoogd:2008id,vanderSchee:2013pia} is neglected.
In order to model event-to-event flutuations, we model the entropy density at $\tau_0$ using the \trento{} Monte Carlo generator~\cite{Moreland:2014oya} with the $p=0$ prescription (corresponding to an entropy density proportional to $\sqrt{T_AT_B}$, where $T_A$ and $T_B$ are the thickness functions of incoming nuclei~\cite{Miller:2007ri}), which has been employed successfully in phenomenological applications~\cite{Giacalone:2017dud}.
The entropy density profile is normalized so that the multiplicity per unit rapidity in central collisions matches the value extracted from experimental data~\cite{Hanus:2019fnc}. 
After a thermalization time $\tau_0=0.4$~fm/c~\cite{Kolb:2000fha}, we evolve this initial condition through 2+1 dimensional boost-invariant ideal hydrodynamics using the MUSIC code~\cite{Schenke:2010nt,Schenke:2011bn,Paquet:2015lta} with a realistic equation of state inspired by lattice QCD~\cite{Huovinen:2009yb}.
Freeze out is done at $T_f=135$~MeV as discussed in Sec.~\ref{s:freezeout}.
We do not implement Cooper--Frye freeze out. 
Instead, we directly obtain $\Omega(u)$ from Eq.~(\ref{probau}), where the integration runs over the freeze-out hypersurface. 
We evolve $\sim 30-40$  different initial conditions in each centrality bin and average $\Omega(u)$ over events. 

\begin{figure}[ht]
\begin{center}
\includegraphics[width=\linewidth]{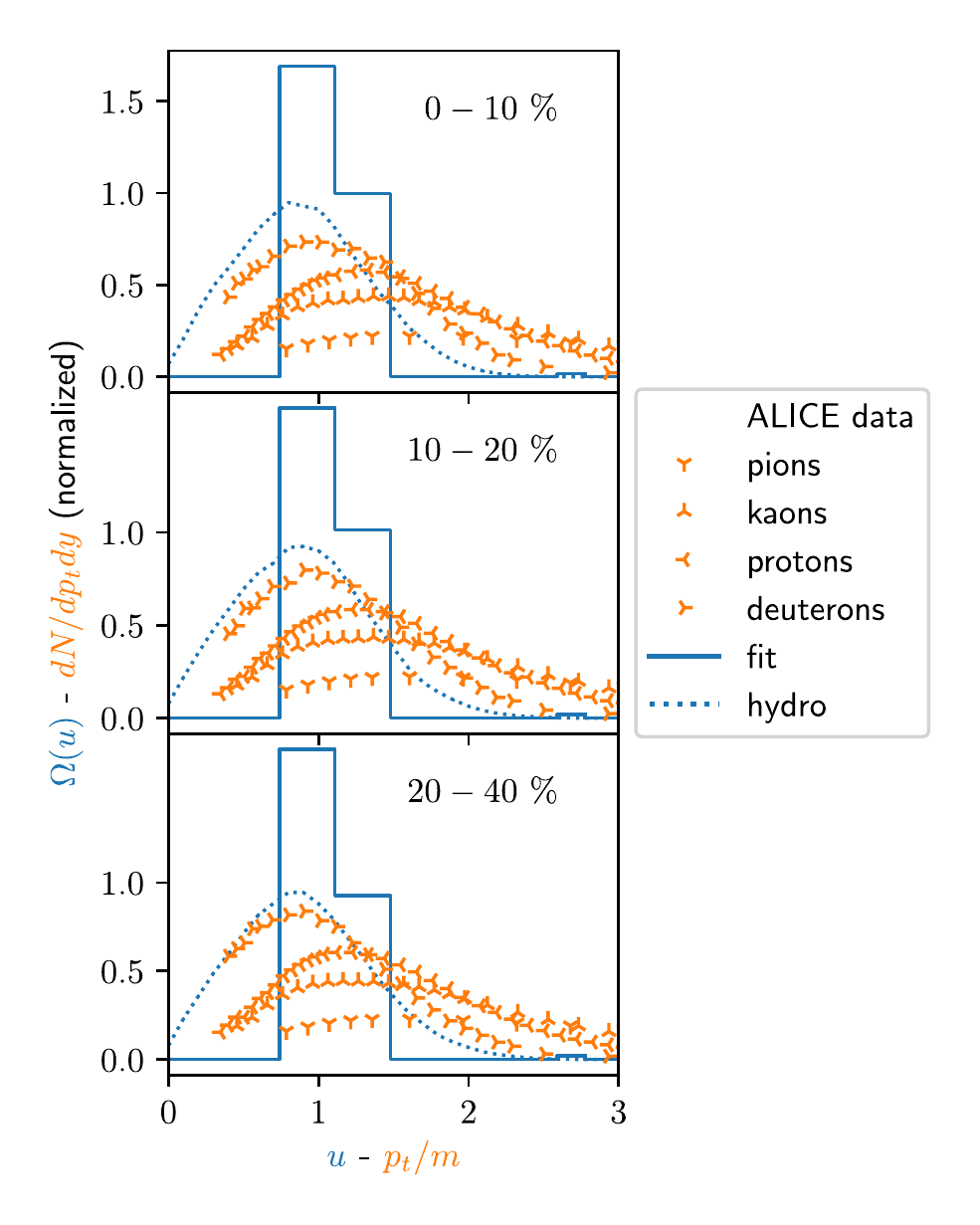} 
\end{center}
\caption{(Color online)
  Symbols: Normalized distributions of $p_t/m$ for pions, kaons, protons~\cite{Adam:2015kca} and deuterons~\cite{Acharya:2017dmc} in Pb+Pb collisions at $\sqrt{s_{\rm NN}}=2.76$~TeV in three centrality windows: 0-10\%, 10-20\%, 20-40\%. 
  Solid lines: normalized distribution of the fluid transverse velocity $u$ resulting from a combined fit of these spectra using Eq.~(\ref{boostedthermal}). 
 Dotted lines:  normalized distribution of $u$ from our event-by-event hydrodynamic calculation. 
 It is calculated in both cases using $p(u)\equiv\Omega(u)/\int_0^{\infty} \Omega(u')du'$. 
}
\label{fig:velocityspectra}
\end{figure}

The maximum of $\Omega(u)$ is at $u\simeq 1$ in the hydrodynamic calculation, as in the distribution extracted from data. 
The effective volume $\int\Omega(u)du$ is also comparable, 
which is a consequence of the fact that the hydrodynamic model predicts the correct multiplicity.
The generalized blast-wave fit returns a  distribution $\Omega(u)$ which is much narrower than that from the hydrodynamic calculation. 
We have seen in Sec.~\ref{s:fittohydro} that this increased narrowness is a generic consequence of approximations underlying the blast-wave fit. 
The effect is however much more pronounced than one would expect on this basis alone. (Specifically, $\Omega(u)$ from the fit is narrower in Fig.~\ref{fig:velocityspectra} than in the upper right panel of Fig.~\ref{fig:fittohydro}.)

It is instructive to understand qualitatively how the distribution of the fluid velocity, $\Omega(u)$, relates to the $p_t$ spectra. 
The transverse momentum of a particle emitted at  freeze-out results from the superposition of the collective motion and the random thermal motion~\cite{Ollitrault:2007du}, $p_t=mu+p_{\rm th}$, where $p_{\rm th}$ is the thermal momentum. 
The higher the mass, the smaller the thermal component relative to the collective one.
Therefore, one expects the distribution of $u$ to approach the distribution of $p_t/m$ for massive particles~\cite{Borghini:2005kd}.
Fig.~\ref{fig:velocityspectra} presents the normalized probability distribution of $p_t/m$ for pions, kaons, protons, deuterons, together with the discrete distribution of the fluid velocity extracted from a combined fit to these spectra, and with the continuous distribution of the fluid velocity from our hydrodynamic calculation. 
Note that deuterons were not included in Figs.~\ref{fig:identifiedspectra} and \ref{fig:Omegaidentified}, because we wanted to make use of the finer centrality binning which is available for the three lighter species. 
Deuterons are useful here, because their velocity distribution comes closer to that of the fluid velocity $u$, due to their much higher mass. 
This is clearly seen when comparing the deuteron spectrum with the smooth distribution of $u$ from the hydrodynamic calculation. 
The distribution $\Omega(u)$ returned by the fit is, however, very different. 
This is in part explained by the coarse binning in $u$ used for the fit, and also by the generic difference between hydrodynamics and blast-wave fits observed in Fig.~\ref{fig:fittohydro}. 
Still, there is a sizable deuteron yield at low $p_t/m$, while $\Omega(u)$ from the fit is identically zero for $u<0.8$.
We do not have a good explanation for this difference.

\begin{figure}[ht]
\begin{center}
\includegraphics[scale=1.]{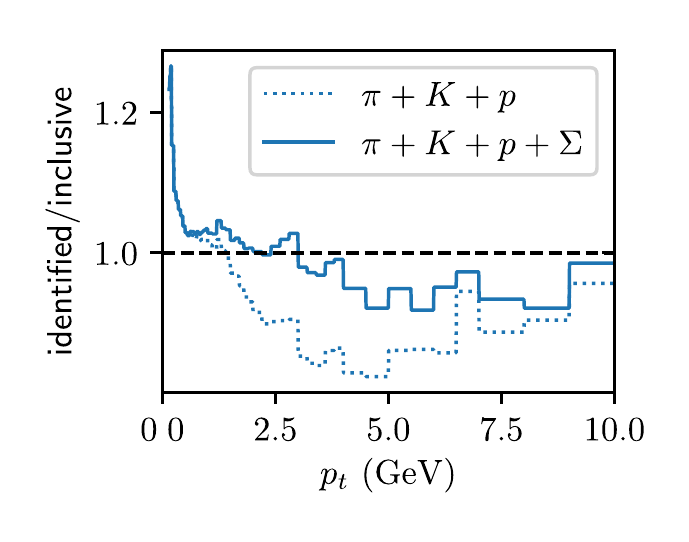} 
\end{center}
\caption{(Color online)
Ratio of the sum of identified particles to all charged particles~\cite{Acharya:2018qsh} as a function of transverse momentum $p_t$, for Pb+Pb collisions at $\sqrt{s_{\rm NN}}=2.76$~TeV in the 0-5\% centrality window. 
The dotted line includes pions, kaons and (anti)protons~\cite{Adam:2015kca}, while the full lines include the contribution of $\Sigma^\pm$ hyperons, inferred from the measured $\Lambda$ spectrum~\cite{Abelev:2013xaa} (see text for details). 
}
\label{fig:identifiedvscharged}
\end{figure}
\subsection{Unidentified spectra and the  low-$p_t$ pion excess} 
\label{s:unidentified}

We now investigate whether the distribution of the fluid velocity can be extracted from unidentified charged particle spectra, which are easier to measure, and for which a broader range of data is available~\cite{Aad:2015wga,Acharya:2018qsh}. 
We first carry out a preliminary check on ALICE data. 
We check if the charged particle spectrum~\cite{Acharya:2018qsh} matches with the sum of identified particle spectra~\cite{Adam:2015kca}. 
This is not trivial because the charged particle spectrum contains a significant contribution from $\Sigma^\pm$ hyperons. 
Since $\Sigma^\pm$ hyperons are not identified, their spectrum is not known. 
We estimate it using a procedure similar to that used by ALICE~\cite{Acharya:2018qsh}. 
Instead of just calculating it with our model, which may introduce errors, we estimate it by rescaling the measured spectrum of $\Lambda$ hyperons, which have the same strangeness content and a similar mass, with a weight obtained from the model.
The distribution $dN/dp_t$ of $\Sigma^\pm$ is then evaluated by multiplying $dN/dp_t$ of $\Lambda$, which is measured~\cite{Abelev:2013xaa}, with the ratio $\Sigma^\pm/\Lambda$ evaluated in the statistical model at this value of $p_t$.  
Fig.\ref{fig:identifiedvscharged} shows the ratio of identified to unidentified particles in central Pb+Pb collisions, before and after including the contribution from $\Sigma^\pm$. 
One sees that this contribution significantly improves agreement at intermediate $p_t$. 
After it is included, agreement is at the level of a few percent, except at very low $p_t$. 
The apparent discrepancy at very low $p_t$ is however likely to be explained by systematic errors, which are as large as 15\%~\cite{Adam:2015kca}. 
The conclusion is that the charged particle spectrum is well understood as the sum of identified-particle spectra. 

\begin{figure}[ht]
\begin{center}
\includegraphics[width=\linewidth]{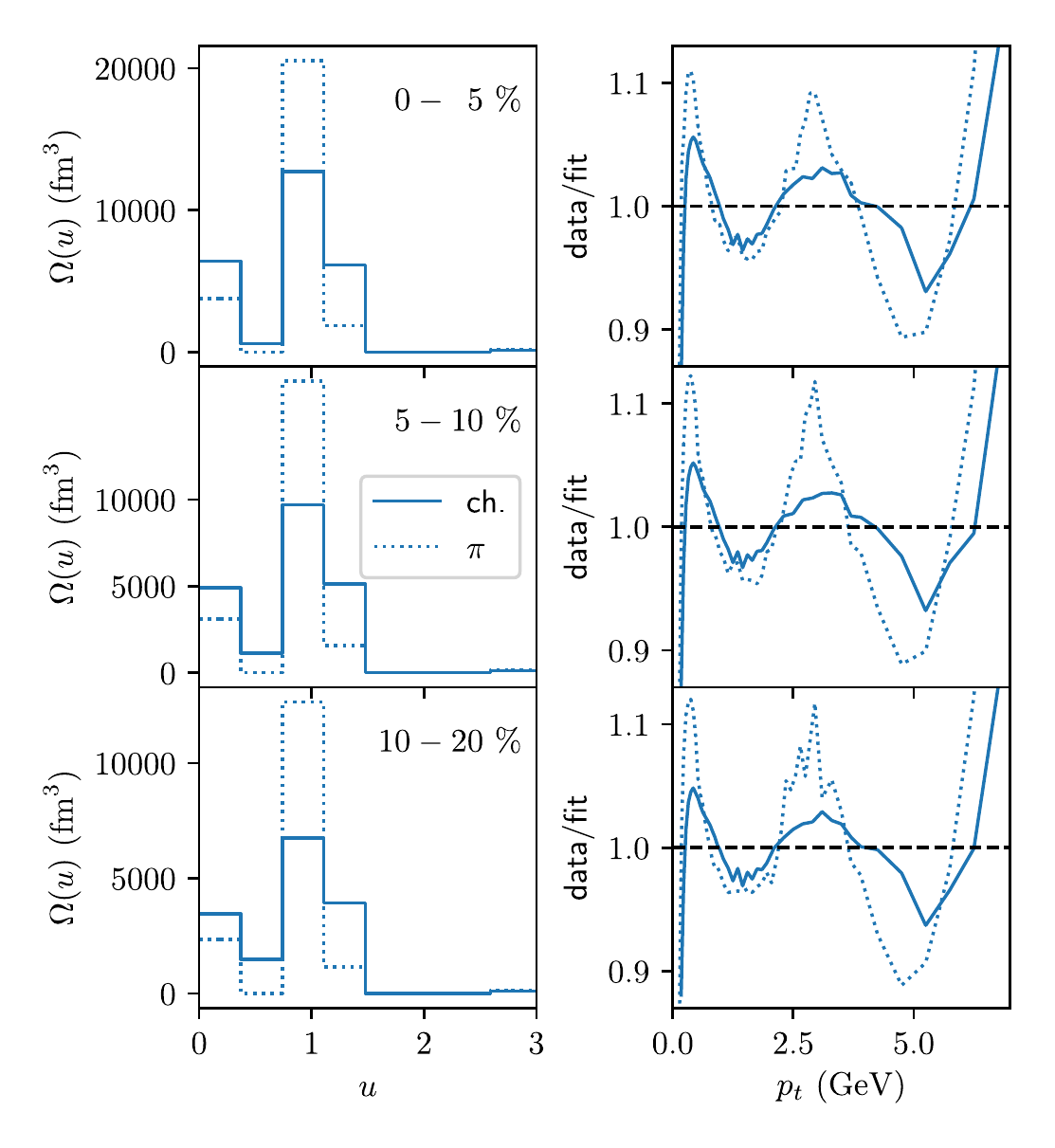} 
\end{center}
\caption{(Color online)
Left: Value of $\Omega(u)$ resulting from the fits to charged particle spectra~\cite{Acharya:2018qsh} (solid lines) and pion spectra~\cite{Adam:2015kca} (dotted lines) in Pb+Pb collisions at $\sqrt{s_{\rm NN}}=2.76$~TeV, using Eq.~(\ref{boostedthermal}) summed over the relevant particle species (see text for details). 
Only three centrality windows are shown. 
Right: Ratio of the measured $dN/dp_t$ divided by the best fit for charged particles (solid lines) and identified pions (dotted lines). 
}
\label{fig:Omegacharged}
\end{figure}

We now extract the distribution of the transverse fluid velocity $\Omega(u)$ from the measured charged particle distribution~\cite{Acharya:2018qsh}. 
We again use Eq.~(\ref{boostedthermal}), where we sum the contributions of pions, kaons, protons, and $\Sigma$ baryons. 
The fit is excellent all the way up to $p_t=6$~GeV/c, as shown in the right panels of Fig.~\ref{fig:Omegacharged}.\footnote{The agreement at high $p_t$ is again due to the small but nonvanishing $\Omega(u)$ in the highest bin $2.6<u<3$, as discussed above for identified particles.}
However, the corresponding values of $\Omega(u)$ differ significantly from those in Fig.~\ref{fig:Omegaidentified} using data on identified particles. 
In addition to the peak around $u\sim 1$, already observed in Fig.~\ref{fig:Omegaidentified}, a second peak appears at $u\sim 0$, corresponding to a fluid at rest. 

We now investigate the origin of the difference between Fig.~\ref{fig:Omegaidentified} and   Fig.~\ref{fig:Omegacharged}.
We have seen (Fig.~\ref{fig:velocityspectra}) that the heaviest particles contain most of the information on $\Omega(u)$. 
Therefore, the combined fit to identified spectra is driven by the heaviest particles used in the fit (protons in the case of Fig.~\ref{fig:identifiedspectra}). 
On the contrary, the fit to the charged particle spectrum is driven by the lightest particles, namely, pions, which represent more than 80\% of the yield (Fig.~\ref{fig:tf}).  
To verify this, we carry out a fit of pion spectra using Eq.~(\ref{boostedthermal}). 
The resulting $\Omega(u)$, shown as a dotted line in Fig.~\ref{fig:Omegacharged}, also presents a peak around $u\sim 0$. 
This peak allows us to achieve good fits also at low $p_t$. 
It is interesting to note that the pion excess at low $p_t$ can be interpreted as coming from a lump of fluid at rest, as if pions at low $p_t$ did not experience the transverse boost imprinted by the pressure. 
This is qualitatively similar to previous interpretations in terms of the formation of a Bose-Einstein condensate of pions~\cite{Begun:2015ifa} or a chiral condensate~\cite{Grossi:2020ezz}, in which soft pions are treated separately from the rest of the fluid~\cite{Grossi:2021gqi}.
Note that a conventional hydrodynamic calculation cannot produce such a peak of $\Omega(u)$ at $u\sim 0$, because $\Omega(u)\propto u$ for small $u$, as seen in 
Fig.~\ref{fig:Omegaidentified}.\footnote{This behavior is easy to understand: 
The three dimensional distribution $\Omega({\bf u})$ in Eq.~(\ref{probauxyz}) is typically continuous at finite at ${\bf u}=0$. 
Upon integration over the azimuthal angle $\phi_u$, the integration measure  $du_xdu_y$ becomes $2\pi udu$, hence the distribution of the transverse fluid velocity is proportional to $u$.} 

\begin{figure}[ht]
\begin{center}
\includegraphics[width=\linewidth]{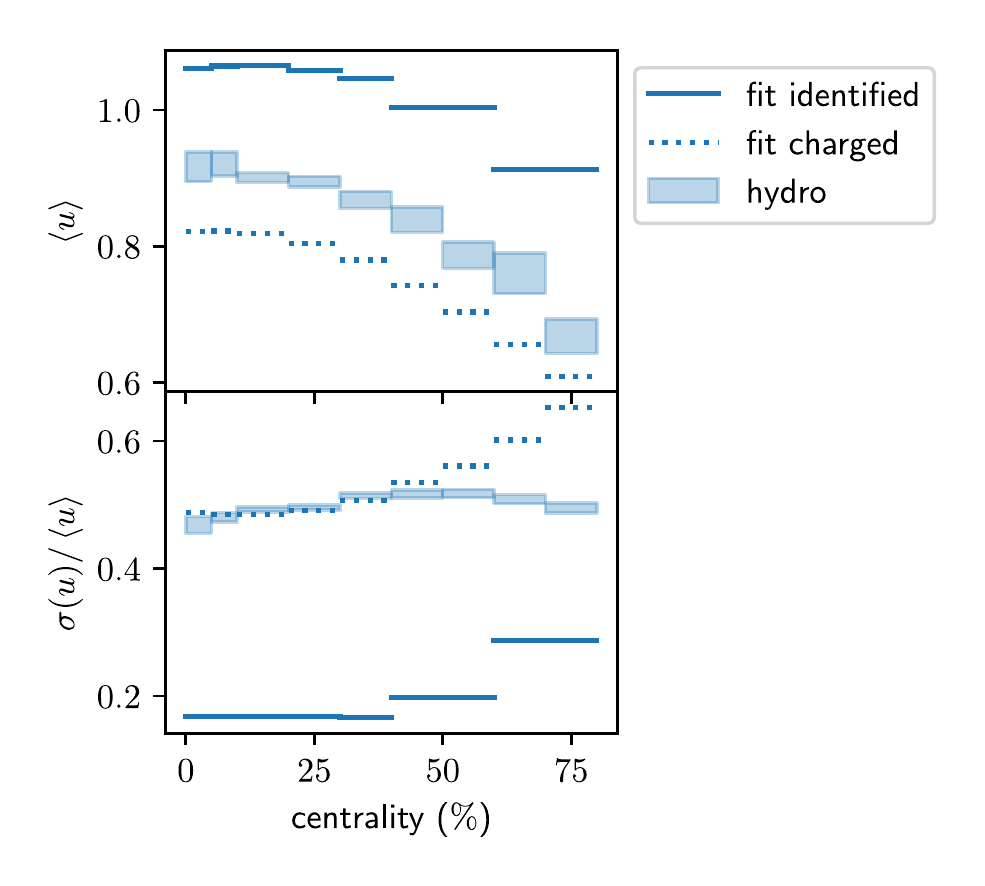} 
\end{center}
\caption{(Color online)
Centrality dependence of the mean value (top) and the relative standard deviation (bottom) of the fluid velocity distribution, as a function of centrality percentile, in Pb+Pb collisions at $\sqrt{s_{\rm NN}}=2.76$~TeV. 
They are evaluated using Eq.~(\ref{defmoments}), where $\Omega(u)$ comes either from the fit to identified spectra (full lines in Fig.~\ref{fig:Omegaidentified}), from the fit to unidentified spectra (full lines in Fig.~\ref{fig:Omegacharged}), or from the hydrodynamic calculation (dotted lines in Fig.~\ref{fig:Omegaidentified}). 
The width of the shaded band used to represent the result of the hydrodynamic calculation is the statistical error due to the finite number of hydrodynamic events, which is evaluated through jackknife resampling. }
\label{fig:centralitydependence}
\end{figure}

\section{Centrality dependence of fluid velocity fluctuations} 
\label{s:centrality}

We finally study the centrality dependence of $p_t$ spectra, and we discuss to what extent it is explained by hydrodynamics. 
Using our generalized blast-wave fit, we interpret the centrality dependence of $p_t$ spectra as stemming from that of the fluid velocity distribution $\Omega(u)$. 
The information contained in $\Omega(u)$ can be expressed in terms of its moments, defined by: 
\begin{equation}
\label{defmoments}
\langle u^n\rangle\equiv\frac{\int_0^\infty u^n\Omega(u) du}{\int_0^\infty \Omega(u) du}. 
\end{equation}
The mean fluid velocity $\langle u\rangle$ and the standard deviation $\sigma_u\equiv \sqrt{\langle u^2\rangle-\langle u\rangle^2}$ encompass an information roughly equivalent  to the mean value and the standard deviation of $p_t$. 
The advantage of choosing $u$ as the variable, rather than $p_t$, is that comparison with hydrodynamic calculations is straightforward, and also more transparent. 

Fig.~\ref{fig:centralitydependence} presents the centrality dependence of the mean fluid velocity, and of the relative standard deviation $\sigma_u/\langle u\rangle$.
The mean value $\langle u\rangle$ extracted from data is smaller if one fits charged spectra than if one fits identifed spectra. 
This is due to the value of $\Omega(u)$ at $u\sim 0$, which is large for charged particles (Fig.~\ref{fig:Omegacharged}), and zero for identified particles  (Fig.~\ref{fig:Omegaidentified}). 
The centrality dependence, however, is similar: the mean velocity mildly decreases as a function of centrality percentile. 
Our event-by-event hydrodynamic calculation returns a value of $\langle u\rangle$ slightly smaller than that extracted from the fit to identified particle spectra. 
This can be attributed to the general trend observed in Fig.~\ref{fig:fittohydro}, that the blast-wave fit overestimates the fluid velocity. 
The decrease of  $\langle u\rangle$ as a function of centrality percentile observed in the data is quantitatively reproduced by our event-by-event hydrodynamic calculation. 
This agreement is not surprising, since hydrodynamics is known to reproduce reasonably well the mild centrality dependence of the mean transverse momentum~\cite{Gardim:2019xjs}. 

We finally discuss results for the relative standard deviation, shown in the bottom panel of Fig.~\ref{fig:centralitydependence}. 
We have already seen in Fig.~\ref{fig:Omegaidentified} that fits to identified particle spectra return a distribution $\Omega(u)$ which is much narrower than an actual hydrodynamic calculation. 
Narrower implies a smaller $\sigma_u/\langle u\rangle$, which is seen in Fig.~\ref{fig:centralitydependence}. 
The fits to charged particle spectra return values of $\sigma_u/\langle u\rangle$ in surprisingly good agreement with the hydrodynamic calculation up to 50\% centrality, even though the distributions $\Omega(u)$ look different (compare the dotted lines in Fig.~\ref{fig:Omegaidentified}  and the solid lines in Fig.~\ref{fig:Omegacharged}). 
The interesting, meaningful result is the mild centrality dependence of the relative standard deviation.  
Both fits (identified and charged) clearly show that it increases as a function of centrality percentile. 
A similar increase is seen in hydrodynamics, up to 50\% centrality. 
The validity of hydrodynamics is expected to get worse as the centrality percentile increases, so that the discrepancies beyond 50\% centrality are not significant. 
In hydrodynamics, the increase of $\sigma_u/\langle u\rangle$ can be ascribed to initial-state fluctuations, which are relatively larger in smaller systems, that is, in less central collisions. 
Our results show that these initial-state fluctuations, which were originally introduced in order to explain data on elliptic flow~\cite{Miller:2003kd,Andrade:2006yh,Alver:2006wh,Holopainen:2010gz} and triangular flow~\cite{Alver:2010gr}, are also instrumental in explaining the centrality dependence of $p_t$ spectra. 

\section{Conclusions}
We have generalized the traditional blast-wave picture to an arbitrary fluid velocity distribution, 
and shown that it can be formulated as a well-defined approximation to a full hydrodynamic calculation, in which one approximates the particle velocity with the fluid velocity at freeze-out. 
Within this approximation, $dN/dp_t$ only involves the distribution of the fluid velocity at freeze out. 
Hence, this distribution of the fluid velocity can be directly obtained by fitting experimental data on $dN/dp_t$, in a way that generalizes usual blast-wave fits. 
Note that our procedure fully includes the feed-down from resonance decays.  

This generalization of the blast-wave approach allows us to obtain reasonable fits of $p_t$ spectra all the way up to 6~GeV, while usual blast-wave fits or hydrodynamic calculations typically fail above 2~GeV. 
If one carries out a combined fit of identified particle spectra, significant deviations between the fit and the data 
are however observed.  
In particular, experimental data show clear evidence of an excess of pions at low $p_t$, relative to ideal hydrodynamics. 
This excess suggests that the viscous correction to the momentum distribution is large at low momentum, unlike usually assumed in viscous hydrodynamic calculations. 

We find that the most probable value of the transverse fluid velocity is 70\% of the velocity of light in Pb+Pb collisions at $\sqrt{s_{\rm NN}}=2.76$~TeV. 
This value is comparable to that found in state-of-the-art hydrodynamic calculations. 
However, the shape of the fluid velocity distribution extracted from data differs from that obtained in these calculations. 
It is much narrower, which is partly due to the approximations underlying the blast-wave approximation.

We have studied the centrality dependence of $p_t$ spectra in nucleus-nucleus collisions.  
Data on Pb+Pb collisions show that the distribution of the fluid velocity becomes broader as the centrality percentile increases. 
We have shown that this broadening is also present in hydrodynamic calculations, where it naturally arises as a consequence of initial-state fluctuations.

\section*{Acknowledgements}
We thank Giuliano Giacalone for discussions and for help with hydrodynamic calculations.
We thank Aleksas Mazeliauskas for help with the implementation of the FastReso code, and for suggesting the work presented in Sec.~\ref{s:fittohydro}.


\begin{thebibliography}{99}

%\cite{Busza:2018rrf}
\bibitem{Busza:2018rrf}
W.~Busza, K.~Rajagopal and W.~van der Schee,
%``Heavy Ion Collisions: The Big Picture, and the Big Questions,''
Ann. Rev. Nucl. Part. Sci. \textbf{68}, 339-376 (2018)
doi:10.1146/annurev-nucl-101917-020852
[arXiv:1802.04801 [hep-ph]].
%217 citations counted in INSPIRE as of 05 May 2021

%\cite{Gardim:2019xjs}
\bibitem{Gardim:2019xjs}
F.~G.~Gardim, G.~Giacalone, M.~Luzum and J.~Y.~Ollitrault,
%``Thermodynamics of hot strong-interaction matter from ultrarelativistic nuclear collisions,''
Nature Phys. \textbf{16}, no.6, 615-619 (2020)
doi:10.1038/s41567-020-0846-4
[arXiv:1908.09728 [nucl-th]].
%27 citations counted in INSPIRE as of 05 May 2021

%\cite{Schnedermann:1993ws}
\bibitem{Schnedermann:1993ws}
E.~Schnedermann, J.~Sollfrank and U.~W.~Heinz,
%``Thermal phenomenology of hadrons from 200-A/GeV S+S collisions,''
Phys. Rev. C \textbf{48}, 2462-2475 (1993)
doi:10.1103/PhysRevC.48.2462
[arXiv:nucl-th/9307020 [nucl-th]].
%991 citations counted in INSPIRE as of 05 May 2021

%\cite{Ollitrault:1992bk}
\bibitem{Ollitrault:1992bk}
J.~Y.~Ollitrault,
%``Anisotropy as a signature of transverse collective flow,''
Phys. Rev. D \textbf{46}, 229-245 (1992)
doi:10.1103/PhysRevD.46.229
%1293 citations counted in INSPIRE as of 05 May 2021

%\cite{Ackermann:2000tr}
\bibitem{Ackermann:2000tr}
K.~H.~Ackermann \textit{et al.} [STAR],
%``Elliptic flow in Au + Au collisions at (S(NN))**(1/2) = 130 GeV,''
Phys. Rev. Lett. \textbf{86}, 402-407 (2001)
doi:10.1103/PhysRevLett.86.402
[arXiv:nucl-ex/0009011 [nucl-ex]].
%764 citations counted in INSPIRE as of 03 May 2021

%\cite{Alver:2010gr}
\bibitem{Alver:2010gr}
B.~Alver and G.~Roland,
%``Collision geometry fluctuations and triangular flow in heavy-ion collisions,''
Phys. Rev. C \textbf{81}, 054905 (2010)
[erratum: Phys. Rev. C \textbf{82}, 039903 (2010)]
doi:10.1103/PhysRevC.82.039903
[arXiv:1003.0194 [nucl-th]].
%799 citations counted in INSPIRE as of 01 May 2021

%\cite{Siemens:1978pb}
\bibitem{Siemens:1978pb}
P.~J.~Siemens and J.~O.~Rasmussen,
%``Evidence for a blast wave from compress nuclear matter,''
Phys. Rev. Lett. \textbf{42}, 880-887 (1979)
doi:10.1103/PhysRevLett.42.880
%351 citations counted in INSPIRE as of 01 May 2021

%\cite{Abelev:2008ab}
\bibitem{Abelev:2008ab}
B.~I.~Abelev \textit{et al.} [STAR],
%``Systematic Measurements of Identified Particle Spectra in $p p, d^+$ Au and Au+Au Collisions from STAR,''
Phys. Rev. C \textbf{79}, 034909 (2009)
doi:10.1103/PhysRevC.79.034909
[arXiv:0808.2041 [nucl-ex]].
%849 citations counted in INSPIRE as of 05 May 2021

%\cite{Abelev:2013vea}
\bibitem{Abelev:2013vea}
B.~Abelev \textit{et al.} [ALICE],
%``Centrality dependence of $\pi$, K, p production in Pb-Pb collisions at $\sqrt{s_{NN}}$ = 2.76 TeV,''
Phys. Rev. C \textbf{88}, 044910 (2013)
doi:10.1103/PhysRevC.88.044910
[arXiv:1303.0737 [hep-ex]].
%606 citations counted in INSPIRE as of 03 May 2021

%\cite{Retiere:2003kf}
\bibitem{Retiere:2003kf}
F.~Retiere and M.~A.~Lisa,
%``Observable implications of geometrical and dynamical aspects of freeze out in heavy ion collisions,''
Phys. Rev. C \textbf{70}, 044907 (2004)
doi:10.1103/PhysRevC.70.044907
[arXiv:nucl-th/0312024 [nucl-th]].
%313 citations counted in INSPIRE as of 01 May 2021

%\cite{Kurkela:2018wud}
\bibitem{Kurkela:2018wud}
A.~Kurkela, A.~Mazeliauskas, J.~F.~Paquet, S.~Schlichting and D.~Teaney,
%``Matching the Nonequilibrium Initial Stage of Heavy Ion Collisions to Hydrodynamics with QCD Kinetic Theory,''
Phys. Rev. Lett. \textbf{122}, no.12, 122302 (2019)
doi:10.1103/PhysRevLett.122.122302
[arXiv:1805.01604 [hep-ph]].
%83 citations counted in INSPIRE as of 07 May 2021

%\cite{Cooper:1974mv}
\bibitem{Cooper:1974mv}
F.~Cooper and G.~Frye,
%``Comment on the Single Particle Distribution in the Hydrodynamic and Statistical Thermodynamic Models of Multiparticle Production,''
Phys. Rev. D \textbf{10}, 186 (1974)
doi:10.1103/PhysRevD.10.186
%975 citations counted in INSPIRE as of 01 May 2021

%\cite{Noronha-Hostler:2013gga}
\bibitem{Noronha-Hostler:2013gga}
J.~Noronha-Hostler, G.~S.~Denicol, J.~Noronha, R.~P.~G.~Andrade and F.~Grassi,
%``Bulk Viscosity Effects in Event-by-Event Relativistic Hydrodynamics,''
Phys. Rev. C \textbf{88}, no.4, 044916 (2013)
doi:10.1103/PhysRevC.88.044916
[arXiv:1305.1981 [nucl-th]].
%159 citations counted in INSPIRE as of 04 May 2021

%\cite{Niemi:2015qia}
\bibitem{Niemi:2015qia}
H.~Niemi, K.~J.~Eskola and R.~Paatelainen,
%``Event-by-event fluctuations in a perturbative QCD + saturation + hydrodynamics model: Determining QCD matter shear viscosity in ultrarelativistic heavy-ion collisions,''
Phys. Rev. C \textbf{93}, no.2, 024907 (2016)
doi:10.1103/PhysRevC.93.024907
[arXiv:1505.02677 [hep-ph]].
%240 citations counted in INSPIRE as of 05 May 2021

%\cite{Kanakubo:2019ogh}
\bibitem{Kanakubo:2019ogh}
Y.~Kanakubo, Y.~Tachibana and T.~Hirano,
%``Unified description of hadron yield ratios from dynamical core-corona initialization,''
Phys. Rev. C \textbf{101}, no.2, 024912 (2020)
doi:10.1103/PhysRevC.101.024912
[arXiv:1910.10556 [nucl-th]].
%17 citations counted in INSPIRE as of 01 May 2021

%\cite{Teaney:2001av}
\bibitem{Teaney:2001av}
D.~Teaney, J.~Lauret and E.~V.~Shuryak,
%``A Hydrodynamic description of heavy ion collisions at the SPS and RHIC,''
[arXiv:nucl-th/0110037 [nucl-th]].
%579 citations counted in INSPIRE as of 01 May 2021

%\cite{Petersen:2008dd}
\bibitem{Petersen:2008dd}
H.~Petersen, J.~Steinheimer, G.~Burau, M.~Bleicher and H.~St\"ocker,
%``A Fully Integrated Transport Approach to Heavy Ion Reactions with an Intermediate Hydrodynamic Stage,''
Phys. Rev. C \textbf{78}, 044901 (2008)
doi:10.1103/PhysRevC.78.044901
[arXiv:0806.1695 [nucl-th]].
%338 citations counted in INSPIRE as of 01 May 2021

%\cite{Bernhard:2016tnd}
\bibitem{Bernhard:2016tnd}
J.~E.~Bernhard, J.~S.~Moreland, S.~A.~Bass, J.~Liu and U.~Heinz,
%``Applying Bayesian parameter estimation to relativistic heavy-ion collisions: simultaneous characterization of the initial state and quark-gluon plasma medium,''
Phys. Rev. C \textbf{94}, no.2, 024907 (2016)
doi:10.1103/PhysRevC.94.024907
[arXiv:1605.03954 [nucl-th]].
%290 citations counted in INSPIRE as of 01 May 2021

%\cite{Schenke:2019ruo}
\bibitem{Schenke:2019ruo}
B.~Schenke, C.~Shen and P.~Tribedy,
%``Multi-particle and charge-dependent azimuthal correlations in heavy-ion collisions at the Relativistic Heavy-Ion Collider,''
Phys. Rev. C \textbf{99}, no.4, 044908 (2019)
doi:10.1103/PhysRevC.99.044908
[arXiv:1901.04378 [nucl-th]].
%33 citations counted in INSPIRE as of 01 May 2021

%\cite{Kolb:2003dz}
\bibitem{Kolb:2003dz}
P.~F.~Kolb and U.~W.~Heinz,
%``Hydrodynamic description of ultrarelativistic heavy ion collisions,''
[arXiv:nucl-th/0305084 [nucl-th]].
%889 citations counted in INSPIRE as of 01 May 2021

%\cite{Ollitrault:2007du}
\bibitem{Ollitrault:2007du}
J.~Y.~Ollitrault,
%``Relativistic hydrodynamics for heavy-ion collisions,''
Eur. J. Phys. \textbf{29}, 275-302 (2008)
doi:10.1088/0143-0807/29/2/010
[arXiv:0708.2433 [nucl-th]].
%162 citations counted in INSPIRE as of 01 May 2021

%\cite{Baier:2007ix}
\bibitem{Baier:2007ix}
R.~Baier, P.~Romatschke, D.~T.~Son, A.~O.~Starinets and M.~A.~Stephanov,
%``Relativistic viscous hydrodynamics, conformal invariance, and holography,''
JHEP \textbf{04}, 100 (2008)
doi:10.1088/1126-6708/2008/04/100
[arXiv:0712.2451 [hep-th]].
%760 citations counted in INSPIRE as of 07 May 2021

%\cite{Romatschke:2017ejr}
\bibitem{Romatschke:2017ejr}
P.~Romatschke and U.~Romatschke,
%``Relativistic Fluid Dynamics In and Out of Equilibrium,''
doi:10.1017/9781108651998
[arXiv:1712.05815 [nucl-th]].
%175 citations counted in INSPIRE as of 01 May 2021

%\cite{Heller:2015dha}
\bibitem{Heller:2015dha}
M.~P.~Heller and M.~Spalinski,
%``Hydrodynamics Beyond the Gradient Expansion: Resurgence and Resummation,''
Phys. Rev. Lett. \textbf{115}, no.7, 072501 (2015)
doi:10.1103/PhysRevLett.115.072501
[arXiv:1503.07514 [hep-th]].
%162 citations counted in INSPIRE as of 07 May 2021

%\cite{Romatschke:2017vte}
\bibitem{Romatschke:2017vte}
P.~Romatschke,
%``Relativistic Fluid Dynamics Far From Local Equilibrium,''
Phys. Rev. Lett. \textbf{120}, no.1, 012301 (2018)
doi:10.1103/PhysRevLett.120.012301
[arXiv:1704.08699 [hep-th]].
%118 citations counted in INSPIRE as of 07 May 2021

%\cite{Florkowski:2010cf}
\bibitem{Florkowski:2010cf}
W.~Florkowski and R.~Ryblewski,
%``Highly-anisotropic and strongly-dissipative hydrodynamics for early stages of relativistic heavy-ion collisions,''
Phys. Rev. C \textbf{83}, 034907 (2011)
doi:10.1103/PhysRevC.83.034907
[arXiv:1007.0130 [nucl-th]].
%227 citations counted in INSPIRE as of 01 May 2021

%\cite{Bazow:2013ifa}
\bibitem{Bazow:2013ifa}
D.~Bazow, U.~W.~Heinz and M.~Strickland,
%``Second-order (2+1)-dimensional anisotropic hydrodynamics,''
Phys. Rev. C \textbf{90}, no.5, 054910 (2014)
doi:10.1103/PhysRevC.90.054910
[arXiv:1311.6720 [nucl-th]].
%142 citations counted in INSPIRE as of 01 May 2021

%\cite{Blaizot:2017ucy}
\bibitem{Blaizot:2017ucy}
J.~P.~Blaizot and L.~Yan,
%``Fluid dynamics of out of equilibrium boost invariant plasmas,''
Phys. Lett. B \textbf{780}, 283-286 (2018)
doi:10.1016/j.physletb.2018.02.058
[arXiv:1712.03856 [nucl-th]].
%44 citations counted in INSPIRE as of 01 May 2021

%\cite{Romatschke:2007mq}
\bibitem{Romatschke:2007mq}
P.~Romatschke and U.~Romatschke,
%``Viscosity Information from Relativistic Nuclear Collisions: How Perfect is the Fluid Observed at RHIC?,''
Phys. Rev. Lett. \textbf{99}, 172301 (2007)
doi:10.1103/PhysRevLett.99.172301
[arXiv:0706.1522 [nucl-th]].
%989 citations counted in INSPIRE as of 01 May 2021

%\cite{Heinz:2013th}
\bibitem{Heinz:2013th}
U.~Heinz and R.~Snellings,
%``Collective flow and viscosity in relativistic heavy-ion collisions,''
Ann. Rev. Nucl. Part. Sci. \textbf{63}, 123-151 (2013)
doi:10.1146/annurev-nucl-102212-170540
[arXiv:1301.2826 [nucl-th]].
%849 citations counted in INSPIRE as of 07 May 2021

%\cite{Dusling:2009df}
\bibitem{Dusling:2009df}
K.~Dusling, G.~D.~Moore and D.~Teaney,
%``Radiative energy loss and v(2) spectra for viscous hydrodynamics,''
Phys. Rev. C \textbf{81}, 034907 (2010)
doi:10.1103/PhysRevC.81.034907
[arXiv:0909.0754 [nucl-th]].
%158 citations counted in INSPIRE as of 01 May 2021

%\cite{Molnar:2014fva}
\bibitem{Molnar:2014fva}
D.~Molnar and Z.~Wolff,
%``Self-consistent conversion of a viscous fluid to particles,''
Phys. Rev. C \textbf{95}, no.2, 024903 (2017)
doi:10.1103/PhysRevC.95.024903
[arXiv:1404.7850 [nucl-th]].
%35 citations counted in INSPIRE as of 01 May 2021

%\cite{Borghini:2005kd}
\bibitem{Borghini:2005kd}
N.~Borghini and J.~Y.~Ollitrault,
%``Momentum spectra, anisotropic flow, and ideal fluids,''
Phys. Lett. B \textbf{642}, 227-231 (2006)
doi:10.1016/j.physletb.2006.09.062
[arXiv:nucl-th/0506045 [nucl-th]].
%154 citations counted in INSPIRE as of 01 May 2021

%\cite{Hagedorn:1967tlw}
\bibitem{Hagedorn:1967tlw}
R.~Hagedorn and J.~Ranft,
%``Statistical thermodynamics of strong interactions at high-energies. 2. Momentum spectra of particles produced in pp-collisions,''
Nuovo Cim. Suppl. \textbf{6}, 169-354 (1968)
CERN-TH-851.
%330 citations counted in INSPIRE as of 01 May 2021

%\cite{Bugaev:2002ch}
\bibitem{Bugaev:2002ch}
K.~A.~Bugaev,
%``Relativistic kinetic equations for finite domains and freezeout problem,''
Phys. Rev. Lett. \textbf{90}, 252301 (2003)
doi:10.1103/PhysRevLett.90.252301
[arXiv:nucl-th/0210087 [nucl-th]].
%40 citations counted in INSPIRE as of 01 May 2021

%\cite{Oliinychenko:2014tqa}
\bibitem{Oliinychenko:2014tqa}
D.~Oliinychenko, P.~Huovinen and H.~Petersen,
%``Systematic Investigation of Negative Cooper-Frye Contributions in Heavy Ion Collisions Using Coarse-grained Molecular Dynamics,''
Phys. Rev. C \textbf{91}, no.2, 024906 (2015)
doi:10.1103/PhysRevC.91.024906
[arXiv:1411.3912 [nucl-th]].
%9 citations counted in INSPIRE as of 01 May 2021

%\cite{Mazeliauskas:2019ifr}
\bibitem{Mazeliauskas:2019ifr}
A.~Mazeliauskas and V.~Vislavicius,
%``Temperature and fluid velocity on the freeze-out surface from $\pi$, $K$, $p$ spectra in pp, p-Pb and Pb-Pb collisions,''
Phys. Rev. C \textbf{101}, no.1, 014910 (2020)
doi:10.1103/PhysRevC.101.014910
[arXiv:1907.11059 [hep-ph]].
%14 citations counted in INSPIRE as of 01 May 2021

%\cite{Yang:2020oig}
\bibitem{Yang:2020oig}
Z.~Yang and R.~J.~Fries,
%``Parameterizing Smooth Viscous Fluid Dynamics With a Viscous Blast Wave,''
[arXiv:2007.11777 [nucl-th]].
%2 citations counted in INSPIRE as of 01 May 2021

%\cite{Mazeliauskas:2018irt}
\bibitem{Mazeliauskas:2018irt}
A.~Mazeliauskas, S.~Floerchinger, E.~Grossi and D.~Teaney,
%``Fast resonance decays in nuclear collisions,''
Eur. Phys. J. C \textbf{79}, no.3, 284 (2019)
doi:10.1140/epjc/s10052-019-6791-7
[arXiv:1809.11049 [nucl-th]].
%16 citations counted in INSPIRE as of 01 May 2021

%\cite{Bjorken:1982qr}
\bibitem{Bjorken:1982qr}
J.~D.~Bjorken,
%``Highly Relativistic Nucleus-Nucleus Collisions: The Central Rapidity Region,''
Phys. Rev. D \textbf{27}, 140-151 (1983)
doi:10.1103/PhysRevD.27.140
%3257 citations counted in INSPIRE as of 07 May 2021

%\cite{Alba:2020jir}
\bibitem{Alba:2020jir}
P.~Alba, V.~M.~Sarti, J.~Noronha-Hostler, P.~Parotto, I.~Portillo-Vazquez, C.~Ratti and J.~M.~Stafford,
%``Influence of hadronic resonances on the chemical freeze-out in heavy-ion collisions,''
Phys. Rev. C \textbf{101}, no.5, 054905 (2020)
doi:10.1103/PhysRevC.101.054905
[arXiv:2002.12395 [hep-ph]].
%15 citations counted in INSPIRE as of 05 May 2021

%\cite{Adam:2015kca}
\bibitem{Adam:2015kca}
J.~Adam \textit{et al.} [ALICE],
%``Centrality dependence of the nuclear modification factor of charged pions, kaons, and protons in Pb-Pb collisions at $\sqrt{s_{\rm NN}}=2.76$ TeV,''
Phys. Rev. C \textbf{93}, no.3, 034913 (2016)
doi:10.1103/PhysRevC.93.034913
[arXiv:1506.07287 [nucl-ex]].
%110 citations counted in INSPIRE as of 03 May 2021

%\cite{Broniowski:2001uk}
\bibitem{Broniowski:2001uk}
W.~Broniowski and W.~Florkowski,
%``Strange particle production at RHIC in a single freezeout model,''
Phys. Rev. C \textbf{65}, 064905 (2002)
doi:10.1103/PhysRevC.65.064905
[arXiv:nucl-th/0112043 [nucl-th]].
%128 citations counted in INSPIRE as of 01 May 2021

%\cite{Bebie:1991ij}
\bibitem{Bebie:1991ij}
H.~Bebie, P.~Gerber, J.~L.~Goity and H.~Leutwyler,
%``The Role of the entropy in an expanding hadronic gas,''
Nucl. Phys. B \textbf{378}, 95-128 (1992)
doi:10.1016/0550-3213(92)90005-V
%149 citations counted in INSPIRE as of 01 May 2021

%\cite{Huovinen:2007xh}
\bibitem{Huovinen:2007xh}
P.~Huovinen,
%``Chemical freeze-out temperature in hydrodynamical description of Au+Au collisions at s(NN)**(1/2) = 200-GeV,''
Eur. Phys. J. A \textbf{37}, 121-128 (2008)
doi:10.1140/epja/i2007-10611-3
[arXiv:0710.4379 [nucl-th]].
%68 citations counted in INSPIRE as of 05 May 2021

%\cite{Devetak:2019lsk}
\bibitem{Devetak:2019lsk}
D.~Devetak, A.~Dubla, S.~Floerchinger, E.~Grossi, S.~Masciocchi, A.~Mazeliauskas and I.~Selyuzhenkov,
%``Global fluid fits to identified particle transverse momentum spectra from heavy-ion collisions at the Large Hadron Collider,''
JHEP \textbf{06}, 044 (2020)
doi:10.1007/JHEP06(2020)044
[arXiv:1909.10485 [hep-ph]].
%18 citations counted in INSPIRE as of 05 May 2021

%\cite{Vovchenko:2018fmh}
\bibitem{Vovchenko:2018fmh}
V.~Vovchenko, M.~I.~Gorenstein and H.~Stoecker,
%``Finite resonance widths influence the thermal-model description of hadron yields,''
Phys. Rev. C \textbf{98}, no.3, 034906 (2018)
doi:10.1103/PhysRevC.98.034906
[arXiv:1807.02079 [nucl-th]].
%26 citations counted in INSPIRE as of 01 May 2021

%\cite{Andronic:2018qqt}
\bibitem{Andronic:2018qqt}
A.~Andronic, P.~Braun-Munzinger, B.~Friman, P.~M.~Lo, K.~Redlich and J.~Stachel,
%``The thermal proton yield anomaly in Pb-Pb collisions at the LHC and its resolution,''
Phys. Lett. B \textbf{792}, 304-309 (2019)
doi:10.1016/j.physletb.2019.03.052
[arXiv:1808.03102 [hep-ph]].
%37 citations counted in INSPIRE as of 01 May 2021

%\cite{Andronic:2017pug}
\bibitem{Andronic:2017pug}
A.~Andronic, P.~Braun-Munzinger, K.~Redlich and J.~Stachel,
%``Decoding the phase structure of QCD via particle production at high energy,''
Nature \textbf{561}, no.7723, 321-330 (2018)
doi:10.1038/s41586-018-0491-6
[arXiv:1710.09425 [nucl-th]].
%324 citations counted in INSPIRE as of 04 May 2021

%\cite{Ryu:2017qzn}
\bibitem{Ryu:2017qzn}
S.~Ryu, J.~F.~Paquet, C.~Shen, G.~Denicol, B.~Schenke, S.~Jeon and C.~Gale,
%``Effects of bulk viscosity and hadronic rescattering in heavy ion collisions at energies available at the BNL Relativistic Heavy Ion Collider and at the CERN Large Hadron Collider,''
Phys. Rev. C \textbf{97}, no.3, 034910 (2018)
doi:10.1103/PhysRevC.97.034910
[arXiv:1704.04216 [nucl-th]].
%65 citations counted in INSPIRE as of 01 May 2021

%\cite{Dubla:2018czx}
\bibitem{Dubla:2018czx}
A.~Dubla, S.~Masciocchi, J.~M.~Pawlowski, B.~Schenke, C.~Shen and J.~Stachel,
%``Towards QCD-assisted hydrodynamics for heavy-ion collision phenomenology,''
Nucl. Phys. A \textbf{979}, 251-264 (2018)
doi:10.1016/j.nuclphysa.2018.09.046
[arXiv:1805.02985 [nucl-th]].
%23 citations counted in INSPIRE as of 01 May 2021

%\cite{Everett:2020xug}
\bibitem{Everett:2020xug}
D.~Everett \textit{et al.} [JETSCAPE],
%``Multi-system Bayesian constraints on the transport coefficients of QCD matter,''
[arXiv:2011.01430 [hep-ph]].
%28 citations counted in INSPIRE as of 05 May 2021

%\cite{Melo:2019mpn}
\bibitem{Melo:2019mpn}
I.~Melo and B.~Tom\'a\v{s}ik,
%``Kinetic freeze-out in central heavy-ion collisions between 7.7 and 2760 GeV per nucleon pair,''
J. Phys. G \textbf{47}, no.4, 045107 (2020)
doi:10.1088/1361-6471/ab5f03
[arXiv:1908.03023 [nucl-th]].
%9 citations counted in INSPIRE as of 01 May 2021

%\cite{Tang:2008ud}
\bibitem{Tang:2008ud}
Z.~Tang, Y.~Xu, L.~Ruan, G.~van Buren, F.~Wang and Z.~Xu,
%``Spectra and radial flow at RHIC with Tsallis statistics in a Blast-Wave description,''
Phys. Rev. C \textbf{79}, 051901 (2009)
doi:10.1103/PhysRevC.79.051901
[arXiv:0812.1609 [nucl-ex]].
%133 citations counted in INSPIRE as of 05 May 2021

%\cite{Alqahtani:2017tnq}
\bibitem{Alqahtani:2017tnq}
M.~Alqahtani, M.~Nopoush, R.~Ryblewski and M.~Strickland,
%``Anisotropic hydrodynamic modeling of 2.76 TeV Pb-Pb collisions,''
Phys. Rev. C \textbf{96}, no.4, 044910 (2017)
doi:10.1103/PhysRevC.96.044910
[arXiv:1705.10191 [nucl-th]].
%48 citations counted in INSPIRE as of 01 May 2021

%\cite{Teaney:2003kp}
\bibitem{Teaney:2003kp}
D.~Teaney,
%``The Effects of viscosity on spectra, elliptic flow, and HBT radii,''
Phys. Rev. C \textbf{68}, 034913 (2003)
doi:10.1103/PhysRevC.68.034913
[arXiv:nucl-th/0301099 [nucl-th]].
%751 citations counted in INSPIRE as of 01 May 2021

%\cite{Gasser:1983yg}
\bibitem{Gasser:1983yg}
J.~Gasser and H.~Leutwyler,
%``Chiral Perturbation Theory to One Loop,''
Annals Phys. \textbf{158}, 142 (1984)
doi:10.1016/0003-4916(84)90242-2
%4364 citations counted in INSPIRE as of 06 May 2021

%\cite{Colangelo:2001df}
\bibitem{Colangelo:2001df}
G.~Colangelo, J.~Gasser and H.~Leutwyler,
%``$\pi \pi$ scattering,''
Nucl. Phys. B \textbf{603}, 125-179 (2001)
doi:10.1016/S0550-3213(01)00147-X
[arXiv:hep-ph/0103088 [hep-ph]].
%951 citations counted in INSPIRE as of 01 May 2021

%\cite{Grossi:2020ezz}
\bibitem{Grossi:2020ezz}
E.~Grossi, A.~Soloviev, D.~Teaney and F.~Yan,
%``Transport and hydrodynamics in the chiral limit,''
Phys. Rev. D \textbf{102}, no.1, 014042 (2020)
doi:10.1103/PhysRevD.102.014042
[arXiv:2005.02885 [hep-th]].
%5 citations counted in INSPIRE as of 01 May 2021

%\cite{Grossi:2021gqi}
\bibitem{Grossi:2021gqi}
E.~Grossi, A.~Soloviev, D.~Teaney and F.~Yan,
%``Soft pions and transport near the chiral critical point,''
[arXiv:2101.10847 [nucl-th]].
%3 citations counted in INSPIRE as of 01 May 2021

%\cite{Aamodt:2010cz}
\bibitem{Aamodt:2010cz}
K.~Aamodt \textit{et al.} [ALICE],
%``Centrality dependence of the charged-particle multiplicity density at mid-rapidity in Pb-Pb collisions at $\sqrt{s_{NN}}=2.76$ TeV,''
Phys. Rev. Lett. \textbf{106}, 032301 (2011)
doi:10.1103/PhysRevLett.106.032301
[arXiv:1012.1657 [nucl-ex]].
%695 citations counted in INSPIRE as of 01 May 2021

%\cite{Paatelainen:2013eea}
\bibitem{Paatelainen:2013eea}
R.~Paatelainen, K.~J.~Eskola, H.~Niemi and K.~Tuominen,
%``Fluid dynamics with saturated minijet initial conditions in ultrarelativistic heavy-ion collisions,''
Phys. Lett. B \textbf{731}, 126-130 (2014)
doi:10.1016/j.physletb.2014.02.018
[arXiv:1310.3105 [hep-ph]].
%39 citations counted in INSPIRE as of 01 May 2021

%\cite{Vredevoogd:2008id}
\bibitem{Vredevoogd:2008id}
J.~Vredevoogd and S.~Pratt,
%``Universal Flow in the First Stage of Relativistic Heavy Ion Collisions,''
Phys. Rev. C \textbf{79}, 044915 (2009)
doi:10.1103/PhysRevC.79.044915
[arXiv:0810.4325 [nucl-th]].
%90 citations counted in INSPIRE as of 01 May 2021

%\cite{vanderSchee:2013pia}
\bibitem{vanderSchee:2013pia}
W.~van der Schee, P.~Romatschke and S.~Pratt,
%``Fully Dynamical Simulation of Central Nuclear Collisions,''
Phys. Rev. Lett. \textbf{111}, no.22, 222302 (2013)
doi:10.1103/PhysRevLett.111.222302
[arXiv:1307.2539 [nucl-th]].
%146 citations counted in INSPIRE as of 01 May 2021

%\cite{Moreland:2014oya}
\bibitem{Moreland:2014oya}
J.~S.~Moreland, J.~E.~Bernhard and S.~A.~Bass,
%``Alternative ansatz to wounded nucleon and binary collision scaling in high-energy nuclear collisions,''
Phys. Rev. C \textbf{92}, no.1, 011901 (2015)
doi:10.1103/PhysRevC.92.011901
[arXiv:1412.4708 [nucl-th]].
%204 citations counted in INSPIRE as of 01 May 2021

%\cite{Miller:2007ri}
\bibitem{Miller:2007ri}
M.~L.~Miller, K.~Reygers, S.~J.~Sanders and P.~Steinberg,
%``Glauber modeling in high energy nuclear collisions,''
Ann. Rev. Nucl. Part. Sci. \textbf{57}, 205-243 (2007)
doi:10.1146/annurev.nucl.57.090506.123020
[arXiv:nucl-ex/0701025 [nucl-ex]].
%1445 citations counted in INSPIRE as of 03 May 2021

%\cite{Giacalone:2017dud}
\bibitem{Giacalone:2017dud}
G.~Giacalone, J.~Noronha-Hostler, M.~Luzum and J.~Y.~Ollitrault,
%``Hydrodynamic predictions for 5.44 TeV Xe+Xe collisions,''
Phys. Rev. C \textbf{97}, no.3, 034904 (2018)
doi:10.1103/PhysRevC.97.034904
[arXiv:1711.08499 [nucl-th]].
%75 citations counted in INSPIRE as of 05 May 2021

%\cite{Hanus:2019fnc}
\bibitem{Hanus:2019fnc}
P.~Hanus, A.~Mazeliauskas and K.~Reygers,
%``Entropy production in pp and Pb-Pb collisions at energies available at the CERN Large Hadron Collider,''
Phys. Rev. C \textbf{100}, no.6, 064903 (2019)
doi:10.1103/PhysRevC.100.064903
[arXiv:1908.02792 [hep-ph]].
%13 citations counted in INSPIRE as of 01 May 2021

%\cite{Kolb:2000fha}
\bibitem{Kolb:2000fha}
P.~F.~Kolb, P.~Huovinen, U.~W.~Heinz and H.~Heiselberg,
%``Elliptic flow at SPS and RHIC: From kinetic transport to hydrodynamics,''
Phys. Lett. B \textbf{500}, 232-240 (2001)
doi:10.1016/S0370-2693(01)00079-X
[arXiv:hep-ph/0012137 [hep-ph]].
%466 citations counted in INSPIRE as of 01 May 2021

%\cite{Schenke:2010nt}
\bibitem{Schenke:2010nt}
B.~Schenke, S.~Jeon and C.~Gale,
%``(3+1)D hydrodynamic simulation of relativistic heavy-ion collisions,''
Phys. Rev. C \textbf{82}, 014903 (2010)
doi:10.1103/PhysRevC.82.014903
[arXiv:1004.1408 [hep-ph]].
%304 citations counted in INSPIRE as of 04 May 2021

%\cite{Schenke:2011bn}
\bibitem{Schenke:2011bn}
B.~Schenke, S.~Jeon and C.~Gale,
%``Higher flow harmonics from (3+1)D event-by-event viscous hydrodynamics,''
Phys. Rev. C \textbf{85}, 024901 (2012)
doi:10.1103/PhysRevC.85.024901
[arXiv:1109.6289 [hep-ph]].
%258 citations counted in INSPIRE as of 01 May 2021

%\cite{Paquet:2015lta}
\bibitem{Paquet:2015lta}
J.~F.~Paquet, C.~Shen, G.~S.~Denicol, M.~Luzum, B.~Schenke, S.~Jeon and C.~Gale,
%``Production of photons in relativistic heavy-ion collisions,''
Phys. Rev. C \textbf{93}, no.4, 044906 (2016)
doi:10.1103/PhysRevC.93.044906
[arXiv:1509.06738 [hep-ph]].
%181 citations counted in INSPIRE as of 01 May 2021

%\cite{Huovinen:2009yb}
\bibitem{Huovinen:2009yb}
P.~Huovinen and P.~Petreczky,
%``QCD Equation of State and Hadron Resonance Gas,''
Nucl. Phys. A \textbf{837}, 26-53 (2010)
doi:10.1016/j.nuclphysa.2010.02.015
[arXiv:0912.2541 [hep-ph]].
%492 citations counted in INSPIRE as of 01 May 2021

%\cite{Acharya:2017dmc}
\bibitem{Acharya:2017dmc}
S.~Acharya \textit{et al.} [ALICE],
%``Measurement of deuteron spectra and elliptic flow in Pb\textendash{}Pb collisions at $\sqrt{s_{\mathrm {NN}}}$ = 2.76 TeV at the LHC,''
Eur. Phys. J. C \textbf{77}, no.10, 658 (2017)
doi:10.1140/epjc/s10052-017-5222-x
[arXiv:1707.07304 [nucl-ex]].
%28 citations counted in INSPIRE as of 03 May 2021

%\cite{Acharya:2018qsh}
\bibitem{Acharya:2018qsh}
S.~Acharya \textit{et al.} [ALICE],
%``Transverse momentum spectra and nuclear modification factors of charged particles in pp, p-Pb and Pb-Pb collisions at the LHC,''
JHEP \textbf{11}, 013 (2018)
doi:10.1007/JHEP11(2018)013
[arXiv:1802.09145 [nucl-ex]].
%118 citations counted in INSPIRE as of 06 May 2021

%\cite{Abelev:2013xaa}
\bibitem{Abelev:2013xaa}
B.~B.~Abelev \textit{et al.} [ALICE],
%``$K^0_S$ and $\Lambda$ production in Pb-Pb collisions at $\sqrt{s_{NN}}$ = 2.76 TeV,''
Phys. Rev. Lett. \textbf{111}, 222301 (2013)
doi:10.1103/PhysRevLett.111.222301
[arXiv:1307.5530 [nucl-ex]].
%329 citations counted in INSPIRE as of 03 May 2021

%\cite{Aad:2015wga}
\bibitem{Aad:2015wga}
G.~Aad \textit{et al.} [ATLAS],
%``Measurement of charged-particle spectra in Pb+Pb collisions at $\sqrt{{s}_\mathsf{{NN}}} = 2.76$ TeV with the ATLAS detector at the LHC,''
JHEP \textbf{09}, 050 (2015)
doi:10.1007/JHEP09(2015)050
[arXiv:1504.04337 [hep-ex]].
%152 citations counted in INSPIRE as of 05 May 2021

%\cite{Begun:2015ifa}
\bibitem{Begun:2015ifa}
V.~Begun and W.~Florkowski,
%``Bose-Einstein condensation of pions in heavy-ion collisions at the CERN Large Hadron Collider (LHC) energies,''
Phys. Rev. C \textbf{91}, 054909 (2015)
doi:10.1103/PhysRevC.91.054909
[arXiv:1503.04040 [nucl-th]].
%44 citations counted in INSPIRE as of 01 May 2021

%\cite{Miller:2003kd}
\bibitem{Miller:2003kd}
M.~Miller and R.~Snellings,
%``Eccentricity fluctuations and its possible effect on elliptic flow measurements,''
[arXiv:nucl-ex/0312008 [nucl-ex]].
%199 citations counted in INSPIRE as of 01 May 2021

%\cite{Andrade:2006yh}
\bibitem{Andrade:2006yh}
R.~Andrade, F.~Grassi, Y.~Hama, T.~Kodama and O.~Socolowski, Jr.,
%``On the necessity to include event-by-event fluctuations in experimental evaluation of elliptical flow,''
Phys. Rev. Lett. \textbf{97}, 202302 (2006)
doi:10.1103/PhysRevLett.97.202302
[arXiv:nucl-th/0608067 [nucl-th]].
%159 citations counted in INSPIRE as of 01 May 2021

%\cite{Alver:2006wh}
\bibitem{Alver:2006wh}
B.~Alver \textit{et al.} [PHOBOS],
%``System size, energy, pseudorapidity, and centrality dependence of elliptic flow,''
Phys. Rev. Lett. \textbf{98}, 242302 (2007)
doi:10.1103/PhysRevLett.98.242302
[arXiv:nucl-ex/0610037 [nucl-ex]].
%360 citations counted in INSPIRE as of 01 May 2021

%\cite{Holopainen:2010gz}
\bibitem{Holopainen:2010gz}
H.~Holopainen, H.~Niemi and K.~J.~Eskola,
%``Event-by-event hydrodynamics and elliptic flow from fluctuating initial state,''
Phys. Rev. C \textbf{83}, 034901 (2011)
doi:10.1103/PhysRevC.83.034901
[arXiv:1007.0368 [hep-ph]].
%194 citations counted in INSPIRE as of 01 May 2021
\end{thebibliography}
\end{document}